\documentclass[twocolumn]{aastex63}

\usepackage{hyperref}
\usepackage{datetime}
\usepackage{natbib}
\usepackage{graphicx}
\usepackage[T1]{fontenc}
\usepackage[utf8]{inputenc}
\usepackage[english]{babel}
\usepackage{color}
\usepackage{float}
\usepackage{amsfonts}
\usepackage[fleqn]{amsmath}
\usepackage{pgfplots}
\usepackage{multirow}
\usepackage{booktabs}
\usetikzlibrary{spy}
\usepackage{bm}
\usepackage{stackengine}
\usepackage{parskip} 
\usepackage[percent]{overpic}
\usepackage{aas_macros}

\usepackage[inline,ignoremode]{trackchanges}    

\addeditor{MO}
\addeditor{MH}
\addeditor{DW}
\newcommand{\der}[3]{\frac{d^{#3} {#1}}{d {#2}^{#3}}}

\newcommand{\pder}[3]{\frac{{\partial}^{#3} {#1}}{{\partial} {#2}^{#3}}}
\newcommand{\ind}[1]{\textrm{\scriptsize #1} }
\newcommand{\limr}[1]{_{\mathrm{#1}}} 
\newcommand{\logba}[2]{\log_\textup{#2}{#1}}

\newcommand{\cm}{\centi\meter}
\newcommand{\G}{\gauss}
\newcommand{\muG}{\mu G}
\newcommand{\yt}{{\texttt{yt}}}
\newcommand{\pc}{{\mathrm{pc}}}
\newcommand{\kpc}{{\mathrm{kpc}}}

\newcommand{\Myr}{{\mathrm{Myr}}}
\renewcommand{\cm}{{\mathrm{cm}}}
\newcommand{\s}{{\mathrm{s}}}
\newcommand{\erg}{{\mathrm{erg}}}
\newcommand{\Msun}{M_{\odot}}
\newcommand{\CR}{{\mathrm{CR}}}
\newcommand{\eCR}{{e_\mathrm{CR}}}
\newcommand{\PCR}{{P_\mathrm{CR}}}
\newcommand{\SCR}{{S_\mathrm{CR}}}
\newcommand{\h}{1/2}
\renewcommand{\th}{3/2}
\renewcommand{\l}{{l}}
\newcommand{\lmh}{{l-\h}}
\newcommand{\lph}{{l+\h}}
\newcommand{\lmth}{{l-\th}}
\newcommand{\lmo}{{\l-1}}
\newcommand{\lpo}{{\l+1}}

\renewcommand{\L}{{\mathrm{L}}}
\newcommand{\LR}{{\mathrm{LR}}}
\newcommand{\R}{{\mathrm{R}}}

\newcommand{\QLR}{Q^{\mathrm{LR}}}
\newcommand{\SLR}{S^{\mathrm{LR}}}
\newcommand{\nLR}{{n^{\textrm{LR}}}}
\newcommand{\eLR}{{e^{\textrm{LR}}}}
\newcommand{\lo}{\mathrm{lo}}
\newcommand{\up}{\mathrm{up}}
\newcommand{\llo}{l_{\lo}}
\newcommand{\lup}{l_{\up}}
\newcommand{\ql}{{q_{\l}}}
\newcommand{\qlmo}{{q_{\lmo}}}

\newcommand{\flmh}{f_\lmh}

\newcommand{\flmth}{{f_{\lmth}}}
\newcommand{\plmh}{p_\lmh}
\newcommand{\plph}{p_\lph}
\newcommand{\nl}{{n_{\l}}}
\newcommand{\el}{{e_{\l}}}
\newcommand{\n}{{\mathrm{n}}}
\newcommand{\e}{{\mathrm{e}}}
\newcommand{\ups}{{\mathrm{u}}}
\newcommand{\pu}{p_\ups}
\newcommand{\pulmh}{{p_{\ups,\lmh}} }
\newcommand{\pulmht}{p^3_{\ups,\lmh} }
\newcommand{\pulmhf}{p^4_{\ups,\lmh} }
\newcommand{\dnulmh}{dn_{\ups,\lmh } (\Delta t)}
\newcommand{\dnulph}{dn_{\ups,\lph } (\Delta t)}
\newcommand{\deulmh}{de_{\ups,\lmh } (\Delta t)}
\newcommand{\deulph}{de_{\ups,\lph } (\Delta t)}
\newcommand{\Rl}{{R_{\l}}}
\newcommand{\kappaLRn}{\kappa^{\mathrm{LRn}}}

\newcommand{\kparal}{\kappa_\parallel}
\newcommand{\kperp}{\kappa_\perp}
\newcommand{\pdl}{\tilde{p}} 
\newcommand{\p}{\pdl}
\newcommand{\nbin}{N_{\mathrm{bin}}}
\newcommand{\pfix}{p_{\mathrm{fix}}}

\newcommand{\pL}{p_\L}
\newcommand{\pR}{p_\R}
\newcommand{\plo}{p_{\mathrm{lo}}}
\newcommand{\pup}{p_{\mathrm{up}}}
\newcommand{\esmall}{\epsilon_{\mathrm{small}}}
\newcommand{\logten}{\log_\mathrm{10}}

\newcommand{\cflcrs}{\mathrm{C}_\mathrm{CRspectr}}
\newcommand{\pdllo}{\pdl_{\mathrm{lo}}}
\newcommand{\pdlup}{\pdl_{\mathrm{up}}}
\newcommand{\pdlbrr}{\pdl\limr{br}^\R}
\newcommand{\pdlbrl}{\pdl\limr{br}^\L}
\newcommand{\fdl}{\tilde{f}}
\newcommand{\edlsmall}{\tilde{\epsilon}\limr{small}}
\newcommand{\creeff}{{\epsilon_{\mathrm{CRe}}}}
\newcommand{\crpeff}{{\epsilon_{\mathrm{CRp}}}}
\newcommand{\funit}{f_{\ind{unit}}}
\newcommand{\epsunit}{\epsilon_{\ind{unit}}}
\newcommand{\ubr}{u\limr{br}}
\received{...}
\accepted{...}
\published{...}
\submitjournal{The Astrophysical Journal Supplement Series}
   \usepackage{silence}
\begin{document}
\title{Implementation of CR Energy SPectrum (CRESP) algorithm  in PIERNIK MHD code.\\
I. Spectrally resolved propagation of CR electrons on Eulerian grids.} 
\shorttitle{CRESP algorithm of PIERNIK MHD code.}
\author[0000-0003-4810-2244]{Mateusz A. {Ogrodnik}}
\affiliation{Institute of Astronomy, Faculty of Physics, Astronomy and Informatics, Nicolaus Copernicus University,\\
ul. Grudziadzka 5/7, PL-87-100 Toruń, Poland}
\author[0000-0002-2370-5631]{Michał Hanasz}
\affiliation{Institute of Astronomy, Faculty of Physics, Astronomy and Informatics, Nicolaus Copernicus University,\\ 
ul. Grudziadzka 5/7, PL-87-100 Toruń, Poland}
\author[0000-0003-3054-6461]{Dominik Wóltański}
\affiliation{Institute of Astronomy, Faculty of Physics, Astronomy and Informatics, Nicolaus Copernicus University,\\ 
ul. Grudziadzka 5/7, PL-87-100 Toruń, Poland}

\shortauthors{M. A. Ogrodnik, M. Hanasz, D. Wóltański}
\correspondingauthor{Mateusz A. {Ogrodnik}}
\email{mateusz.ogrodnik@astri.umk.pl,\\ mateusz.antoni.ogrodnik@gmail.com}
\date{today}
\setlength{\abovedisplayskip}{2.5pt}
\setlength{\belowdisplayskip}{2.5pt}
\begin{abstract}

We present an efficient algorithm to follow spectral evolution of Cosmic Rays (CR) coupled with an MHD system on Eulerian grids. 
The algorithm is designed for studies of CR energy spectrum evolution in MHD simulations of a galactic interstellar medium.
The base algorithm for CR transport relies on the two-moment piece-wise power-law method, known also as Coarse Grained
Momentum Final Volume (CGMV), for solving the Fokker-Planck CR transport equation, with a low number of momentum-bins
 extending over several decades of the momentum coordinate. 
{ We propose an extension of the CGMV with a novel feature which allows momentum boundaries to change in response to CR momentum gains or losses near the extremes of the population distribution.}
Our extension involves a special treatment of momentum bins containing spectral cutoff.
Contrary to the regular bins of fixed width, those bins have variable-width, and their outer edges coincide with spectral cutoffs. 
The cutoff positions are estimated from the particle number density and energy density in the outer bins for an assumed 
small value of an additional parameter representing the smallest physically significant level of CR spectral energy density. 

We performed a series of elementary tests to validate the algorithm and demonstrated, whenever possible,  
that  results of the test simulations correspond, with a reasonable accuracy, to the results of analogous analytical solutions. 
In a more complex test of galactic CR-driven wind problem we obtained results consistent with expectations regarding 
the effects of advection, diffusion, adiabatic, and synchrotron cooling of a CR population.
\end{abstract}


\section{Introduction}

Cosmic rays take part in many astrophysical phenomena: cosmic protons and nuclei influence dynamical as well as chemical evolution of galaxies \citep{Grenier2015},
e.g., quenching the process of stellar formation \citep{2018NatAs...2...83T} and enriching chemical composition of the ISM via spallations. CR electrons and other leptons interact with magnetic and radiation fields, thus being directly linked to the observations of synchrotron radiation \citep{10.1111/j.1365-2966.2011.19114.x, Strong2011}, which they produce at the cost of its kinetic energy.
Observations show that radio emission in disk galaxies varies in extent at different wavelengths, while spectral indices also indicate spatial variation \citep{2011MNRAS.412.2396F, 2018A&A...615A..98M} in galaxies seen: face-on (harder spectral indices in spiral arms) and edge-on (harder spectral index in the galactic plane). The CR spectrum varies subtly in space and time, at least in our local environment, which is observed locally in e.g., solar modulation of low energy CRs \citep{2013LRSP...10....3P} and measurements of Voyager spacecrafts \citep{Cummings_2016} as well as inferred from observations of distant galaxies.

Recent numerical studies combining CR propagation with magnetohydrodynamical evolution of the interstellar medium (ISM) treated CRs as a massless relativistic fluid  in the two-fluid system, described through the momentum-integrated CR diffusion-advection equation. These investigations  focus on 
three major topics: (1) modeling Parker instability in the ISM due to buoyancy of CRs  \citep[e.g.,][]{2003A&A...412..331H,2004apj...607..828k,2018apj...860...97h} which leads to the formation of magnetic loops and evacuation of CRs  from the galactic disk, (2) amplification of large-scale galactic magnetic fields by the 'cosmic ray-driven dynamo', originally proposed by \cite{1992ApJ...401..137P}:
\cite{2004ApJ...605L..33H,2009A&A...498..335H,2009ApJ...706L.155H,2011ApJ...733L..18K,2015a&a...575a..93k,2010a&a...510a..97s,2014A&A...562A.136S,2018A&A...611A...7S}, and (3)  acceleration of powerful galactic winds \citep[see, e.g.,][]{2012mnras.423.2374u,2013ApJ...777L..38H,boothetal2013,salembryan2014,2016ApJ...816L..19G,2018MNRAS.479.3042G,farberetal2018,hopkinsetal2020b}.

Radioobservations of galactic synchrotron emissions provide useful information on the  strength and structure of galactic magnetic fields \citep[see][for a comprehensive review]{2015A&ARv..24....4B}. However, a unique reconstruction of a three-dimensional magnetic field structure from the observational data is not an easy task. On the other hand, theoretical models and numerical simulations provide insights into physical processes connecting magnetic field evolution to the complex dynamics of interstellar gas and cosmic rays, but they offer a limited access to useful observables, such as the spectrum of synchrotron radio emission.
These circumstances highlight the need for time-dependent spectral modeling of the CR electron population if one needs to confront MHD models of galactic ISM against observations of real galaxies.

We note the efforts to interpret observational data with the help of spectrally resolved CR propagation codes, such as GALPROP \citep{Moskalenko&Strong1998}, USINE \citep{2011a&a...526a.101p}, PICCARD \citep{2013arxiv1308.2829w},  DRAGON \citep{2014phrvd..89h3007g}, and SPINNAKER \citep{2018MNRAS.476..158H}, which provide numerical solutions for CR diffusion and advection in fixed magnetic fields and stationary background flows.
We emphasize that our strategy, presented in this paper, is different: we solve the CR transport equations together with the system of MHD equations to obtain time-dependent solutions for gas, magnetic fields, and the energy spectrum of CRs.
We present Cosmic Ray Energy SPectrum (CRESP) -- a new numerical algorithm designed to simulate momentum-dependent  diffusion-advection propagation of CR electrons, coupled with MHD simulations of the thermal component on 3D Eulerian grids. The presented algorithm has been incorporated into the framework of PIERNIK MHD code.

The plan of the paper is as follows: in Section~\ref{sect:approach} we formulate our novel method to solve the problem of CR-electron spectral evolution on Eulerian grids, and in Section \ref{sect:initial_condition} we describe initial conditions.  
In Section \ref{sect:cresp_tests} we describe validation tests of  CRESP algorithm operating in a one-zone mode, and in Section \ref{sect:piernik_tests} we present validation tests of CRESP operating together with the CR spatial transport scheme of PIERNIK.
In Section \ref{sect:piernik_crwind_test} we present test simulations of CR-electron spectrum evolution in a quasi-realistic 3D model of galactic wind and, in Section \ref{sect:conclusions} we conclude our paper.
\section{Numerical algorithms for CR transport equation} \label{sect:approach}
\subsection{Cosmic Ray transport equation}
 The time evolution of distribution function, describing ultra relativistic CR component, can be characterized by simplified diffusion-advection equation \citep{1975MNRAS.172..557S, 1987PhR...154....1B, 2001CoPhC.141...17M}, which can be expressed as
 \begin{eqnarray}
      \pder{f}{t}{} = -\bm{v} \cdot \nabla f + \nabla ( \hat{\kappa} \nabla f) +\frac{1}{3} (\nabla \cdot \bm{v})\,  p \pder{f}{p}{} & \nonumber \\
    + \frac{1}{p^2} \pder{}{p}{} \left[  p^2 b_l(p) f +D_{\rm pp} \pder{f}{p}{}    \right]  + j(p) &
      \label{eq:cr_transp_df}
\end{eqnarray}
where $f = f(\bm{x},p,t)$ is the isotropic part of the distribution function of a population of CR particles, $\bm{v}$ is the velocity of thermal plasma, $\hat{\kappa}$ is the diffusion tensor representing diffusive propagation of CRs in space, and $D_{\rm  pp}$ is the momentum diffusion coefficient.
Similarly, as \citep{2020MNRAS.491..993G}, we ignore for simplicity CR streaming process \citep[see, e.g.,][]{Sharma_2010,2018ApJ...854....5J,2019MNRAS.485.2977T}. We also ignore the second-order Fermi acceleration by setting $D_{\rm pp} = 0$.
The  first two terms on the right-hand side represent advection and diffusion in physical space. The third term represents the adiabatic cooling or heating of CR population, depending on the sign of $\nabla \cdot \bm{v}$. Radiative losses $b_l(\bm{x},p)$ with 2nd order Fermi mechanism are accounted for in the fourth term and, CR sources are represented by
$j = j(\bm{x}, p, t)$. Equation~(\ref{eq:cr_transp_df}) describes temporal evolution of a population of energetic particles subject to strong scattering, which ensures particle distribution remains isotropic on scales much greater than particle scattering-length \citep{1999ApJ...512..105J}.

We are aiming to solve numerically this equation together with the full set of MHD equations describing temporal evolution of the thermal plasma in MHD approximation.  The CR transport equation~(\ref{eq:cr_transp_df}) is coupled with the MHD equations via velocity field $\bm{v}$ (in Lagrangian derivative of the distribution function $f$ and adiabatic cooling/heating term) and magnetic field $\bm{B}$, if the synchrotron cooling process is considered. In a general case CRs provide an additional pressure contribution to the system. In this case the equation of gas motion should include the gradient of the total CR pressure $\nabla p\limr{CR}$, corresponding to the pressure CRs exert on thermal gas. However, the pressure from CR electron population might be ignored due to their relatively low energy density, estimated at about 1\%  of CR proton energy density \citep{2007ARNPS..57..285S}.

\subsection{The two-moment  piece-wise power-law approximation}

Starting with the papers by \citet{1987MNRAS.225..399F} and \citet{1987MNRAS.225..615B},  various numerical methods have been proposed to solve the CR propagation equation (\ref{eq:cr_transp_df}) in momentum space.
\cite{1991MNRAS.249..439K} solved numerically the CR transport equation using a standard  discretization in momentum space assuming piece-wise constant values of distribution function in momentum bins. \cite{2019MNRAS.488.2235W} demonstrated that this kind of discretization requires approximately 50--100 bins per momentum decade to achieve accurate results of numerical integration of CR transport equations coupled with the system of MHD equations. Such a high spectral resolution, needed due to the steep CR spectra and large dynamical range of $f$, is, however, not feasible in multidimensional fluid-dynamical simulations involving cosmic rays.

An alternative method, based on a piece-wise power-law  representation of the distribution function
\begin{equation}
    f_{} (p) = \flmh \left(\frac{p}{\plmh}\right)^{-\ql} \;\; \mbox{for} \;\; p \in [p_\lmh, p_\lph] \label{eq:dist_fun-ppl}
\end{equation}
has been proposed  by \citet{1999ApJ...512..105J}, where $ f_{\lmh}$ are the distribution function amplitudes, defined on the left edges of momentum bins $\plmh$ and spectral indices $\ql$ are attributed to bin interiors.
The piece-wise power-law spectrum is constructed on  logarithmically spaced momentum bins with fixed bin width equal to
\begin{equation}
 \Delta w_l = \log_{10} (\plmh/\plph) = \log ( p\limr{max} / p\limr{min} )/\nbin,
 \label{logpgrid}
\end{equation}
where $\nbin$ is the number of bins.
This approach involves the number density moment (see Appendix \ref{sect:f_power_law}) of the transport equation (\ref{eq:cr_transp_df}) combined with an additional assumption of continuity of the distribution function. This  approach leads, however, to a numerical instability pointed out by  \citep{2020MNRAS.491..993G}, who show that, if  energy is injected at one part of the spectrum, the continuity assumption enforces changes of the local slope across the entire spectrum and results in artificial oscillations of the initially smooth spectrum between a concave and convex spectrum.
The piece-wise power-law representation of distribution function has two degrees of freedom, which can be bound by taking the number density and energy density moments of the diffusion-advection equation (\ref{eq:cr_transp_df}).
\cite{2001CoPhC.141...17M} proposed and implemented in COSMOCR code a method which relies on the direct conversion between the two pairs of distribution function parameters in each bin: $(\flmh, \ql)$ to the number density $\nl$ and energy density $\el$ moments.
Moreover, in the absence of explicit sources, these two moments are conservative quantities, and therefore they are suitable for an accurate evolution with conservative transport schemes in both the spatial and momentum domains. The method is known also as Coarse--Grained Momentum finite Volume  (CGMV) \citep{2005APh....24...75J}. Details of the method are presented in the Appendix.

The two-moment piece-wise power-law method is particularly useful in the case of strong cooling processes such as  synchrotron and inverse-Compton losses of CR electrons  \citep{2001CoPhC.141...17M} or  Coulomb losses of CR protons \cite{2020MNRAS.491..993G}, which produce cut-offs in the distribution function. With momentum bins spanning a wide range of values, the cutoffs usually fall in the middle of the bin, causing only part of the bin being populated. The cutoffs should be accurately followed in order to extract the populated part of the spectrum, as resolving the distribution function beyond the physical cutoffs leads to numerical problems.

Numerical solutions are available in a practically unlimited range of momentum coordinates only if the strong cooling processes, such as synchrotron or Coulomb losses, are ignored. \citet{2005APh....24...75J} modeled the propagation of CRs across an Eulerian grid, using a fully hydrodynamical description, including advection and diffusion processes in physical and momentum spaces.
They imposed the upper boundary condition by setting the particle number density to zero at the high end of the spectrum. At the low  range of the spectrum the distribution function can be matched to the thermal particle distribution.

\cite{2001CoPhC.141...17M} applied the method for the evolution of CR electron spectrum attributed to Lagrangian particles. Advection of the CR spectra resulted directly from the Lagrangian treatment of the particles, while diffusion of CRs was ignored in this approach. In the absence of diffusive mixing of different CR populations in particle codes, the evolution of spectral cutoffs, attributed to individual particles, can be followed directly via equation~(\ref{eq:btot}), therefore numerical integration of the evolution equations~(\ref{eq:ncr_update}) and (\ref{eq:ecr_update_implicit}) is straightforward.  Following this method  \cite{2009ApJ...696.1142M}, \cite{2017ApJ...850....2Y}, and  \cite{2018ApJ...865..144V}  implemented spectrally resolved CR propagation by means of (mesh-less) Lagrangian macro-particles embedded in a classical or relativistic Eulerian MHD system.
However, exchange of CR spectra between different fluid elements becomes essential in the fully Eulerian treatment of CRs, i.e., when advection and diffusion of CRs are to be taken into account on Eulerian grids.

For very low momenta the timescale related to the Coulomb cooling process becomes shorter than the typical time step of MHD simulations. Similarly, for very high momenta the synchrotron and inverse Compton energy losses proceed on arbitrarily short timescales. \cite{2019MNRAS.488.2235W} presented an efficient post-processing code for Cosmic Ray Electron Spectra that are evolved in Time (CREST) on Lagrangian macro-particles. They divided the overall spectrum into three parts treated with different methods. To calculate efficiently the CR electron spectrum with time steps similar to the MHD time step, they used analytical solutions for low and high momenta together with the fully numerical treatment for intermediate momenta. A similar technique based on the combination of numerical integration and analytical prescription of the Coulomb cooling has been applied by \cite{2020MNRAS.491..993G}, who incorporated the combination of analytical and numerical methods into the method by \cite{2001CoPhC.141...17M}. The use of analytical solution helps to overcome the excessive shortening of the integration time-step whenever the cooling timescale is significantly shorter than the hydrodynamical timescale.

Focusing on the consideration on the synchrotron cooling, we note that the process operates most efficiently in the high momentum range of the spectrum.
{ We note that the synchrotron and inverse Compton losses are represented by similar cooling terms in the CR transport equation (\ref{eq:cr_transp_df}), resulting in momentum losses contributing to terms proportional to $p^2$ in equation (\ref{eq:btot}).
The general case includes both the synchrotron and inverse Compton losses, but the methods presented here are insensitive to which of these two processes is actually dominating. Therefore, without loss of generality, we can discuss only the case of synchrotron losses, keeping in mind that one needs only to add the radiation field energy density to magnetic field energy density to retain full generality of the presented considerations.}
The overall momentum-dependent CR transport algorithm  should be able to determine the position of the upper cutoff and should be equipped with appropriate boundary conditions in momentum space. We note that CR spectra
may have different cutoffs in each spatial grid-cell, because local cooling conditions (magnetic and velocity fields) are generally different. Due to advective and diffusive propagation of the CRs across the spatial grid, different populations of particles mix in every cell. The essential part of the numerical problem is to estimate an effective cut-off for the mixture of different populations inflowing from neighboring cells with different cut-offs.

Our aim is to work out a fully Eulerian algorithm, based on the piece-wise power-law method, designed for momentum-dependent modeling of CR propagation in space. Obvious benefits would be a straightforward treatment of a CR diffusion based on the fluid approach and accurate computation of pressure forces coupling the dynamics of CR population with thermal plasma.
We will concentrate our efforts on describing CR electrons, which are the focus of this paper. However we note that the following calculus can be used for CR protons (and other species) in a similar way.

\subsection{Extension of the formalism to Eulerian grids} \label{sect:ext_to_euler}

\subsubsection{The environment of PIERNIK MHD code}

PIERNIK is a  grid-based MHD code, using conservative numerical schemes, available from 
the public {\tt git} repository linked to the code webpage: \url{http://piernik.umk.pl}. 
The functionality of PIERNIK includes the modeling of multiple fluids: gas, dust, cosmic rays, and their gravitational and electromagnetic interactions. PIERNIK is parallelized on the base of MPI library, and its data I/O communication utilizes parallel HDF5 output.
The MHD algorithm is based on the standard set of resistive MHD equations \citep{piernik1,piernik2}
including  uniform and current dependent resistivity \citep{piernik3} that are solved using the RTVD scheme by \citep{jin-xin-95} and \citep{2003ApJS..149..447P}.  PIERNIK code has been recently equipped with the HLLD Riemann solver by \cite{2005JCoPh.208..315M}  combined with the 
\cite{ref_Dedner_divB_cleaning} algorithm, which may serve as an alternative to the simple and robust but more diffusive RTVD numerical scheme. Other  algorithms added recently to the code include the Adaptive Mesh Refinement (AMR) technique, multigrid (MG) Poisson solver, multigrid diffusion solver, and an N-body particle-mesh solver for large-number point masses representing  stellar and dark matter components of galaxies.

\subsubsection{Spatial transport of CRs}\label{sect:spatial_transport}
The numerical  algorithm of CR spatial transport embedded in PIERNIK is based on formulae (\ref{eq:nLR}) for CR number density and (\ref{eq:eLR}) for CR energy density, respectively. No transfers of CRs between momentum bins are involved in this step.

Inter-cell fluxes along the $x$-axis of CR number density $n$, and energy density $e$ computed at cell faces are governed by
\begin{eqnarray}
   F^{\ind{adv,X}}_{x,i\pm \h,j,k,l}= (v_x X_l)_{i\pm \h,j,k},
\end{eqnarray}
where $X$ represents either $n$ or $e$, $v_x$ is the $x$-component of gas velocity, index $l$ refers to the bin number, and face centering of cell centered quantities relies on an interpolation consistent with the fluid advection algorithm in use. Analogous formulae are adopted for advection in the $y$ and $z$ directions. We note that a CR advection scheme appropriate for shock discontinuities should involve CR number density (and  energy density) normalized to thermal gas density \cite[see, e.g.,][]{2007JCoPh.227..776M}. However here we  focus on the formulation of a CR spectrum evolution algorithm and treat all aspects of the spatial transport in the simple way presented.

The numerical algorithm for the anisotropic, magnetic field-aligned diffusion relies entirely on the method presented by \cite{2003A&A...412..331H}. Components of the anisotropic diffusion tensor are
\begin{equation}
\kappa_{ij}(p) = \kappa_{\perp}(p) \delta_{ij}  + \left( \kappa_{\parallel}(p) -  \kappa_{\perp}(p_l)\right) b_i b_j,
\end{equation}
where $\kappa_{\parallel}(p)$ and $\kappa_{\perp}(p)$ are the parallel and perpendicular, momentum-dependent diffusion coefficients, and $b_i =B_i/|B|$ are components of the unit vectors tangent to magnetic field lines.

To achieve the numerical stability of the anisotropic diffusion algorithm one needs (in addition to the standard CFL timestep limitation, see equation (\ref{eq:dt_max_mhd})) a careful interpolation of magnetic field components as well as CR number density and energy density gradients. With an appropriate use of slope limiters, the algorithm turns out to be robust even in the presence of dynamical coupling between thermal gas and CR fluid. Diffusive fluxes along the $x$-axis computed at cell faces are
\begin{equation}
 F^{\ind{dif,X}}_{x,i\pm \h,j,k,l} = \left(\hat{\kappa}^{\ind{X}}_l(p_l) \nabla n_l\right)_{x,i\pm \h,j,k}, 
\end{equation}
where $\hat{\kappa}^{\ind{X}}_l (p_l) $  
represent momentum-averaged diffusion tensors of CR number density $n$ and energy density $e$ in momentum bin No. $l$. Similar relations, involving variations of indices $j$ and $k$, hold for $y$ and $z$ directions, respectively.

In the discrete representation, the three--dimensional conservation law reads
\begin{equation}
  \hspace{-1pt}\begin{array}{ll}
     X_{i,j,k,l}^{t+ d t}=X_{i,j,k,l}^t
              &  -\frac{\Delta t}{\Delta x}
                 \left(F^{X}_{i+\h,j,k,l}-F^{X}_{i-\h,j,k,l} \right)\\
	      &  -\frac{\Delta t}{\Delta y}
                \left(F^{X}_{i,j+\h,k,l}-F^{X}_{i,j-\h,k,l} \right)       \\
	      &  -\frac{\Delta t}{\Delta z}
                \left(F^{X}_{i,j,k+\h,l}-F^{X}_{i,j,k-\h,l} \right) ,
      \end{array}
      \label{eqn:fluxes}
\end{equation}
where $X$ represents either $n$ or $e$, which are volume-averaged number and energy densities, $X_{i,j,k,l}^t $, $X_{i,j,k,l}^{t+dt} $ denote these quantities in $l$-th bin in the cell ${i,j,k}$ at times $t$ and $t+dt$, respectively,
and $F^{\ind{X}}_{i-\frac{1}{2},j,k,l}$, $F^{\ind{X}}_{i+\frac{1}{2},j,k,l}$ represent the sum of advection and diffusion fluxes through the left and right cell boundaries, in $x$-direction,  while the remaining flux components, indexed with $j$ and $k$, respectively, represent transport in $y$- and $z$-directions.

The advection part of the spatial transport is realized in a way analogous to propagation of a passive tracer. The advection and diffusion steps are executed in an operator split manner.
Advection of all CR bins is a part of the dimensionally split fluid update step consisting of the sequence of sweeps in $x$-$y$-$z$ and $z$-$y$-$x$ directions. There are two options for the choice of diffusion algorithms. The default one is the explicit, dimensionally split solver introduced by \cite{2003A&A...412..331H}, which executes the sequence of $x$-$y$-$z$ or  $z$-$y$-$x$ transports before each sequence of three  advection sweeps. The explicit diffusion timestep is subject to the CFL condition
\begin{equation}
  \Delta t_\mathrm{max} = 0.5 \ C_{\ind{cr-dif}} \
  \frac{\min (\Delta x,\Delta y, \Delta z)^2}{\max ( \kappa_{\parallel}(p_l) + \kappa_{\perp}(p_l))}, \label{eq:dt_max_mhd}
\end{equation}
where $C_{\ind{cr-dif}} \leq 1$ is the Courant number for the CR diffusion process and the maximum of diffusion coefficients over the set of momentum bins is used.
An alternative to the explicit method is an implicit multigrid diffusion solver, which mitigates the CFL condition at some higher computational cost.

The formalism presented in the appendix provides formulae (\ref{eq:kappa_n}) and (\ref{eq:kappa_e}) for the bin-averaged diffusion of number density and energy density, respectively. This involves spatial derivatives of the distribution function, which in practical implementation lead to stability problems of the numerical algorithm and to a significant growth of computational costs \citep{2020MNRAS.491..993G}. To avoid these difficulties we  approximate the diffusion of number and energy densities with their values computed for centers of momentum bins. Equal diffusion coefficients for the particle number density and energy density imply that the ratio of $e_l/n_l$ is preserved in the diffusion step, and therefore  the slope $q_l$ of the distribution function does not change within the bin. This leads to the formation of 'teeth' in the distribution function, which are apparent in the pure diffusion tests, but are usually negligible in the case of combined diffusion as well as advection with adiabatic and synchrotron processes (to be shown in Section~\ref{sect:piernik_tests}). This approximation leads to a stable evolution of the CR population, so we decided to use it in the forthcoming series of tests.
Diffusion coefficients are dependent on momenta attributed to the bin centers and normalized as follows:
\begin{equation}
   \kappa_{\parallel,\perp}(p_l) = {\kappa\limr{10\mathrm{k}}}_{\parallel,\perp} \left( \frac{p_l}{10^4 m_e c} \right)^{\alpha_{\kappa}}.
   \label{eq:diffusion-coefficients-approximate}
\end{equation}
where $\kappa\limr{10\mathrm{k}}$ is the parallel diffusion coefficient at the particle momentum $p=10^4 m_e c$, which is representative of particles contributing to synchrotron radio emission at centimeter wavelengths in magnetic fields on the order of a few $\muG$, typical for the interstellar medium.
The power index $\alpha_\kappa$ scales the momentum dependence of the diffusion coefficient. We assume $\alpha_{\kappa} = 0.5$. Usually CR diffusion is dominant  in the direction of the magnetic field (see, e.g., \citet{1990acr..book.....B, 1999ApJ...520..204G}), hence we usually assume that perpendicular diffusion coefficient $\kperp = 1 \% \, \kparal $.
\subsubsection{The problem to solve}
It is important to remember that, due to the synchrotron cooling rate proportional to $p^2$ all electrons cool down below $p_{cut} \propto B^{-2}\, t^{-1}$
in a finite time \citep{1962SvA.....6..317K}, therefore the whole range of momentum space above $p_{cut}$ becomes empty.
Because the synchrotron cooling rate depends on the magnetic field strength, which in turn depends on space coordinates and time, cut-off momenta are different
in different cells of the spatial grid. While CR fluid propagates according to momentum-dependent diffusion-advection equation, the flux of particles through cell boundaries changes the content of momentum bins.
In the presence of spatial transport of CRs, the cut-off momenta of an individual cell depend not only on the local physical conditions, but also on an inflow of particles from the adjacent cells. The essential problem to solve is therefore to estimate the cut-off momenta in all cells of the Eulerian grid after the spatial transport step, which mixes CR populations of neighboring cells.
\subsubsection{The proposed solution} \label{sect:cutoff_momenta-approx}
Each cell of the spatial grid is attributed with a piece-wise power-law CR spectrum defined on a discrete, logarithmically spaced (according to equation (\ref{logpgrid})) momentum grid, named hereafter as {\em fixed momentum grid}.
To solve the problem we need an operational definition  of spectral cut-offs, which is appropriate for the piece-wise representation of CR spectrum. We shall treat similarly the  low- and high-energy parts of the CR spectrum. We denote left and right edges of the spectrum as  $\plo$ and  $\pup$, both dependent on spatial coordinate indexes, which we omit to simplify the notation. The spectrum limits evolve according to the integrated momentum evolution equations (\ref{eq:p_t+dt}).
Variations of the cutoff momenta imply that the  bins of a fixed grid, containing $\plo$ and $\pup$, are filled only partially,
i.e., only the high-energy part of the lowest bin and low-energy part of the highest  bin are filled with particles.  The cut-off momenta $\plo$ and $\pup$  become edges of the outer variable-width bins. We define active bins as those containing CR particles. The remaining bins are inactive.
\begin{figure}
  \hspace{-5pt}\includegraphics[width=0.485\textwidth]{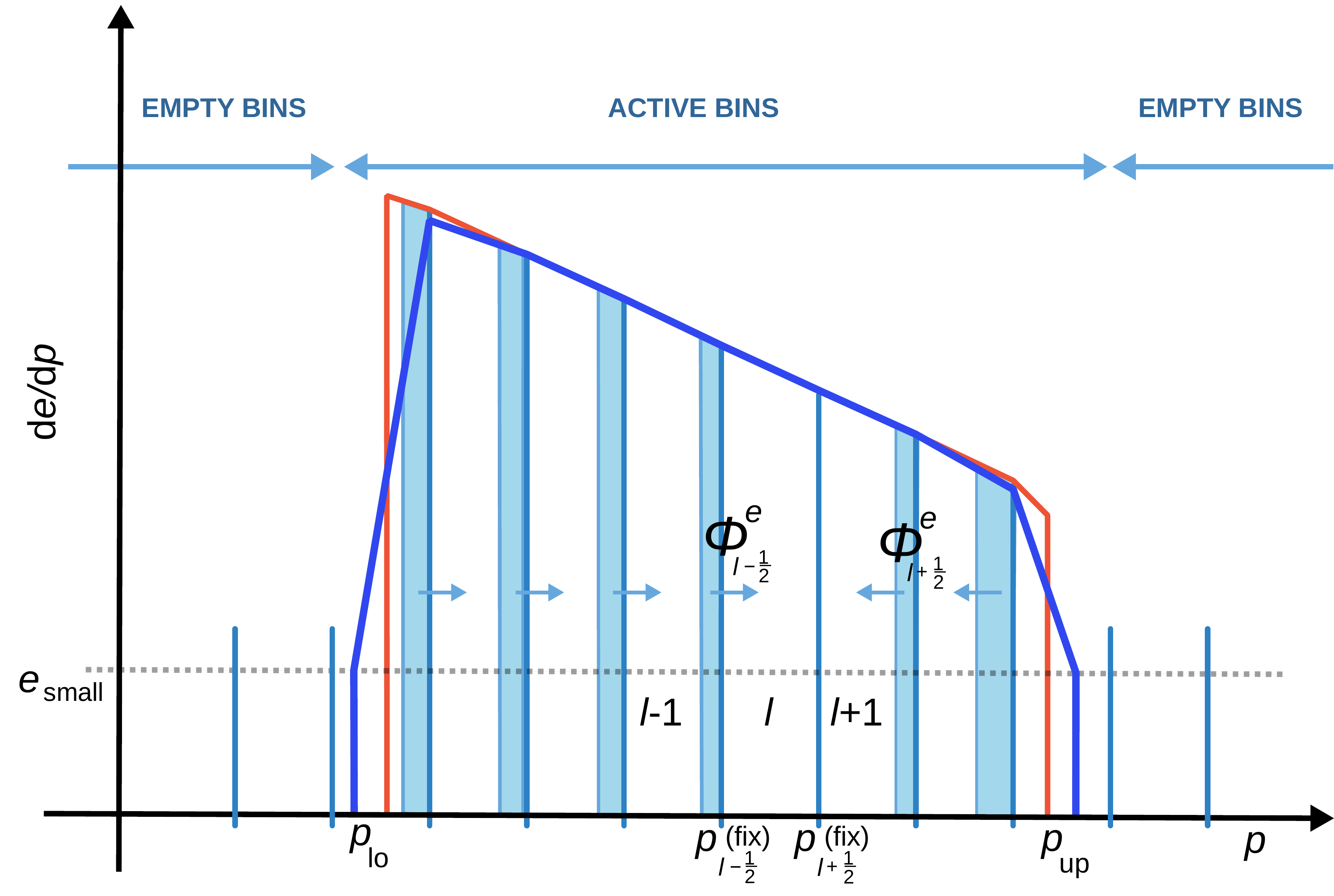}
   \caption{Depiction of domain parametrization in $p$-space within CRESP algorithm, $\der{e}{p}{}$ shown in case of exact (red) and approximated (blue) cutoffs with $\esmall$  limit (gray dashed line). The grid is composed of $\nbin$ sections, encompassed within $\nbin+1$ bin faces. Power-law like spectrum is limited by two cutoff momenta: $p\limr{lo}$, $p\limr{up}$, spanning on fixed momentum grid $\pfix$ (vertical light-blue bars). Net fluxes are marked with $\Phi$ and light-blue areas, indicating compression with intensive cooling in high energy range. Approximating cutoffs results in hardening/steepening slopes of the extreme ends of the spectrum and cutoff momenta values exceeding the exact ones.}\label{fig:example_spectrum_e}
\end{figure}

We shall extract the cut-off information in each cell using the set of known values $(e_{\l}, n_{\l})$ and solving equation~(\ref{eq:enpc-q})
for the unknown $\pup$ or $\plo$, rather than for the slope $q_l$.
We note that in the piece-wise power-law representation  the distribution function values  should be finite at the cut-offs. Otherwise, the slopes  $q_l$ in the outer bins would be infinite.
We need to calculate distribution function values at both edges of the cut-off bin; for one of these edges we can use the solution to equation~(\ref{eq:n_l}) either for $f_{l,L}$ or $f_{l,R}$. That leaves us with one unknown value of $f$, which can be resolved by introducing a free parameter.

We propose the following procedure for the bin reconstruction based on the known values of $e_{\l}$ and $n_{\l}$:
\begin{enumerate}
   \item We define a new parameter $\esmall$ denoting the smallest physically significant value of the spectral energy density
   \begin{equation}
      \epsilon \equiv de/dp = 4 \pi c p^3 f(p).
      \label{eq:spectr_e}
   \end{equation}
   We assume that  $\epsilon = \esmall$ at the outer edge of the bin, where $\esmall$ is a free parameter which will be adopted, with some help of validation tests presented in Section~\ref{sect:cresp_tests}. The values of $p$ at which $\epsilon = \esmall$ are our estimates for $\plo$ and $\pup$.

   \item  Let us denote the number of the bin containing the lower cutoff as $\llo$ and the number of the bin containing the upper cutoff as $\lup$.
   Knowing the values of $(n_{\lup}, e_{\lup})$  together with the left edge  $p_{{\lup},L}$ of the uppermost bin  and $\esmall$ we want to  estimate the upper cutoff momentum $\pup$. Similarly, knowing the values of $(n_{\llo}, e_{\llo})$ together with the right edge of the lowermost bin ${p_{{\llo},R}}$ and $\esmall$ we want to estimate the lower cutoff momentum $\plo$. Graphical presentation of the problem for the lowermost and uppermost  momentum bins is shown in Figure~\ref{fig:example_spectrum_e}.

   \item We define new variables  $ \mu \equiv p_{l,R} / p_{l,L} $ and $ \phi \equiv f_{l,R} / f_{l,L} $ and
   use equation (\ref{eq:dist_fun-eul}) to obtain
   \begin{equation}
      q = -\frac{\logten \phi}{\logten \mu}.
      \label{qphimu}
   \end{equation}

   \item For the lowest active bin $p_{\llo,L } =  \plo$ and $f_{\llo,R} $ are unknowns,   $f_{\llo,L} = \esmall/(4\pi c \plo^3)$  and  $p_{\llo,R} = p_{\llo+\h}$ is the known right (fixed) edge of the lowest bin, and
   \begin{equation}
      \mu_\lo = \frac{p_{\llo+\h}}{\plo}, \qquad  \phi_\lo = \frac{4\pi c \plo^3 f_{\llo,R}}{\esmall},
      \label{mu-phi_lo}
   \end{equation}
   and $q_\lo$ is computed using equation~(\ref{qphimu}).

   For the highest active bin $p_{\lup,R } =  \pup$ and  $f_{\lup,L} $ are unknowns,  $f_{\lup,R} =  \esmall/(4\pi c \pup^3)$  and  $p_{\lup,L} = p_{\lup-\h}$ is the known left (fixed) edge of the highest bin, and
   \begin{equation}
      \mu_\up = \frac{\pup}{p_{\lup-\h}},  \qquad \phi_\up = \frac{\esmall}{4\pi c \pup^3 f_{\lup,L}},
      \label{mu-phi_up}
   \end{equation}
   and $q_\up$ is given by equation~(\ref{qphimu}).

   \item We transform equation~(\ref{eq:enpc-q}) with the aid of equations~(\ref{mu-phi_lo}) to get

   \begin{equation}
      \hspace{-15pt}\frac{e_{\llo}}{n_{\llo} c \, p_{\llo+\h}} = \left\{
      \begin{array}{l l}
         \vspace{6pt}   \frac{1}{\mu_\lo} \frac{\mu_\lo ^ {4 - q_\lo} - 1}{(4 - q_\lo) \log{\mu_\lo}}  & \mathrm{if} \; q_\lo = 3,   \\
         \vspace{6pt}   \frac{1}{\mu_\lo} \frac{(3-q_\lo) \log \mu_\lo }{\mu_\lo ^ {3-q_\lo} -1}       & \mathrm{if} \; q_\lo = 4,  \\
         \frac{1}{\mu_\lo} \frac{3 - q_\lo }{4 - q_\lo}\; \frac{\mu_\lo ^ {4 - q_\lo} - 1 }{ \mu_\lo ^ {3-q_\lo} -1}    & \mathrm{otherwise}.
      \end{array}
      \right.
      \label{eq:e2n_lo}
   \end{equation}
   Formula  (\ref{eq:n_l}) for particle number density $n_{\llo}$ translates to
   \begin{equation}
      \hspace{-15pt}n_{\llo} = \frac{\esmall}{c} \times \left\{
      \begin{array}{l l}
         \vspace{6pt}   \log{\mu_\lo}            & \mathrm{if}\; q_\lo = 3, \\
         \frac{ \mu_\lo^{3 - q_\lo} -1 }{3 - q_\lo } & \mathrm{otherwise}.
      \end{array}
      \right.
      \label{eq:n_lo}
   \end{equation}
   Similarly, for the upper boundary  equation (\ref{eq:enpc-q})  becomes
   \begin{equation}
      \hspace{-0.35cm}\frac{e_{\lup}}{n_{\lup} c \, p_{\lup-\h}} = \left\{
      \begin{array}{l l}
         \vspace{6pt}   \frac{\mu_\up ^ {4 - q_\up} - 1}{(4 - q_\up) \log{\mu_\up}}  & \mathrm{if} \; q_\up = 3   \\
         \vspace{6pt}   \frac{(3-q_\up) \log \mu_\up }{\mu_\up ^ {3-q_\up} -1}       & \mathrm{if} \; q_\up = 4  \\
         \frac{3 - q_\up }{4 - q_\up}\; \frac{\mu_\up ^ {4 - q_\up} - 1 }{ \mu_\up ^ {3-q_\up} -1}    &  \mathrm{otherwise}.
      \end{array}
      \right.
      \label{eq:e2n_up}
   \end{equation}
   Equation~(\ref{eq:n_l}) for $n_{\lup}$  translates to
   \begin{equation}
      \hspace{-10pt}n_{\lup} = \frac{\esmall}{c} \times \left\{
      \begin{array}{l l }
         \frac{\log \mu_\up} { \phi_\up \mu_\up ^3 } & \mathrm{if} \; q_\up = 3 \\
         \frac{1} {\phi_\up \mu_\up ^3 } \frac{\mu_\up ^{3 - q_\up}  - 1} {3 - q_\up} & \mathrm{otherwise}.
      \end{array}
      \right.
      \label{eq:n_up}
   \end{equation}
   \item Solutions to these equations are obtained with the aid of a two-dimensional Newton-Raphson-type method.
   For a given value of l.h.s of the equation systems (\ref{eq:e2n_lo})--(\ref{eq:n_lo}) or  (\ref{eq:e2n_up})--(\ref{eq:n_up}), the numerical procedure finds solutions $(\phi_\lo,\mu_\lo)$ and $(\phi_\up,\mu_\up)$, which are translated into spectrum cutoffs using prescriptions:
   \begin{eqnarray}
      \hspace{-14pt}\plo = &p_{\llo+\h}\frac{1} {\mu_\lo}, \qquad  \hspace{1pt}f_{\llo,R} &= \frac{\esmall} { 4 \pi c \plo^3},
   \end{eqnarray}
   for the lowest bin and
   \begin{eqnarray}
      \pup = &p_{\lup-\h} \, \mu_\up, \quad f_{\lup,L} &= \frac{\esmall} { 4 \pi c \pup^3 } \frac{1}{\phi_\up}
   \end{eqnarray}
   for the highest bin.
\end{enumerate}

\subsection{Structure of the overall algorithm}\label{sect:cresp+mhd}

Having solved the problem of determining spectrum cutoffs on Eulerian grid in Section \ref{sect:cutoff_momenta-approx}, we may proceed to combine the spatial and momentum-space transport schemes to obtain the full framework for CR spectral evolution on the spatial grid.
Momentum-space transport is done by the CRESP module of PIERNIK, in between two sequences of spatial transport sweeps,  $x-y-z$ and $z-y-x$ (see Section \ref{sect:spatial_transport}).
Transport of the CR population over the momentum-grid, executed in all cells of the spatial grid, proceeds in the following order:
\begin{enumerate}
   \item {\em Finding active bins.}
   \begin{enumerate}
      \item We determine the set of bins, where $n_l$ and $e_l$ are non-vanishing and tag the remaining bins empty.

      \item We apply the root-finding method to solve equation (\ref{eq:enpc-q}) for slope $q_l$ in all non-empty bins. In the actual implementation we tabulate solutions of equation (\ref{eq:enpc-q}) at the simulation initialization step and use interpolation methods during the simulation.

      \item Each slope $q_l$ found in the previous step is used to solve equation (\ref{eq:n_l}) for distribution function amplitude $\flmh$.

      \item Using equation  (\ref{eq:dist_fun-eul})) and (\ref{eq:spectr_e}) we compute spectral energy density $\epsilon_\lmh$ and $\epsilon_\lph$ at  left and right edges  $p_\lmh$ and $p_\lph$ of the bin.

      \item We compare $\epsilon_\lmh$ and $\epsilon_\lph$  against $\esmall$ and tag all bins fulfilling the condition
      \begin{equation}
         \max(\epsilon_\lmh,\epsilon_\lph) > \esmall
      \end{equation}
      as active bins. The remaining bins are tagged as inactive for the spectrum evolution algorithm,  regardless of their content, but not for the spatial transport. For the lowermost active bin, associated with the lower cutoff we set $\llo$, while $\lup$ is associated with the uppermost active bin.
   \end{enumerate}
   \item {\em Determining cutoffs.}\\
   We recover the lower cutoff information for the bins containing spectral cutoffs from $n_{\llo}$ and $e_{\llo}$ by solving the system of equation (\ref{mu-phi_lo}), (\ref{eq:e2n_lo}), and (\ref{eq:n_lo}) for the unknowns $\plo$ and $f_{\llo,R}$ ,  as explained in Section \ref{sect:cutoff_momenta-approx}. Similarly, we find the upper cutoffs  from $n_{\lup}$ and $e_{\lup}$ by solving the system of equation (\ref{mu-phi_up}), (\ref{eq:e2n_up}), (\ref{eq:n_up}) for the unknown $\pup$ and $f_{\lup,L}$.
   Using a Newton-Raphson solver we tabulate the solutions at the initialization of the code run and use interpolated values in the time-step integration loop. The values of cutoff momenta  and distribution function amplitudes are now reconstructed in all active bins.

   \item {\em Computing changes of cutoff momenta.}\\
   Particle momentum changes owing to the considered processes are computed for each point of momentum grid via integration of equation (\ref{eq:btot}) in an interval $(t, t+\Delta t)$ resulting in solutions given by equation (\ref{eq:p_t+dt}), which are used to compute changes of cutoff momenta within the timestep. For practical reasons we use a combination of Taylor expansions of these solutions by applying the formulae (\ref{eq:synchrotron_cooling_Taylor3}) and (\ref{eq:adiabatic_cooling_Taylor3}).

   \item {\em Computing particle momentum changes at bin interfaces.}\\
   Upstream momenta $\pu$, defined in equation (\ref{eq:p_u}), are computed in a similar manner and used to determine the direction of the flow in momentum space through bin interfaces. If $\pulmh > \plmh$ the interface is marked as cooling interface, and if $\pulmh < \plmh$ is marked as heating interface.

   \item {\em Computing particle and energy fluxes through bin interfaces}.\\
   Fluxes of particle and energy densities crossing the bin interfaces, $\dnulmh$ and $\deulmh$ (respectively) are computed. For cooling interfaces, if the flow is directed from $l$-th to $(\lmo)$-th bin, equations (\ref{eq:nupw_l}) and (\ref{eq:eupw_l}) are used. Similarly,  for heating interfaces, if the flow is directed from $l$-th to $(\lpo)$-th bin, equations (\ref{eq:nupw_r}) and (\ref{eq:eupw_r}) are used.

   \item {\em Computing source terms for adiabatic changes and radiative losses.}\\
   Gains and losses of particle energy due to interactions of CRs with the thermal plasma and due to radiative losses are included as source terms in equation~(\ref{eq:ecr_update}) and are quantified by the factor $R_l$ defined through equation~(\ref{eq:e_sink_R}) in the case of adiabatic process combined with synchrotron losses.

   \item {\em Numerical integration of particle number density and energy density equations.}\\
   Time-integration of the number density and energy density evolution equations is performed in the first-order operator-split manner. A conservative transport step of $\nl$ and $\el$ in momentum space (due to inter-bin fluxes) is executed in the first step, according to the momentum-advection part of equation (\ref{eq:ncr_update}) and (\ref{eq:ecr_update})
   \begin{eqnarray}
      \hspace{-0.4cm}
      n_l^{t + \Delta t} = & n_l^{t} - \left( \dnulph - \dnulmh \right)\hspace{-2pt},
      \label{eq:ncr_update-1step}\\
      \hspace{-0.4cm}
      \el^{(1)} =& e_l^{t} -  \left( \deulph - \deulmh  \right)\hspace{-2pt}.
      \label{eq:ecr_update-1step}
   \end{eqnarray}
   Energy density is subsequently updated with the source terms
   \begin{equation}
      \el^{t + \Delta t} =  \el^{(1)} \left( 1 -  \Delta t \Rl \right).
      \label{eq:ecr_update-source}
   \end{equation}

\end{enumerate}
The whole procedure is executed in all cells of the spatial grid, where  the values of adiabatic process and synchrotron cooling parameters $u_d$ and $\ubr$ depend on local conditions in individual cells of the spatial grid.
\subsection{Code optimization challenges}\label{sect:challenges}
In order to describe some of the aforementioned processes in detail, one might need to either compute $f$ and $q$ on the run (resulting in an increase of the number of operations) or keep information on these. Our aim was not to store $f$ and $q$ values, but instead keep information on $n$ and $e$, and thus for each computational cell we need $2 \nbin$ fields, still resulting in $N_x \times N_y \times N_z \times 2\nbin$ fields on any grid. Storing information on $f$ and $q$ would double the memory usage by each CR component, posing a dilemma in which we either choose doubling the number of saved quantities or to recompute these quantities, impacting algorithm performance, as obtaining these quantities is computationally expensive.
\citet{Press1996} points out that Newton-Raphson (N-R) convergence is fast, but has poor global convergence properties. For this reason as well as for optimization needs, it is critical to provide the scheme with a good initial guess. This can be achieved by preparing solution maps for each unknown quantity $\mu, \phi$ for both cutoffs in two--dimensional $({e}/{n c p},\; n)$ space. Unknowns first are obtained via linear interpolation of mapped solutions, and then either refined using N-R scheme or used as is, if resolution of maps is satisfactory.

Equation (\ref{eq:enpc-q}) used to obtain $\ql$ can be parametrized with just one term (that is, ${e_l}/{n_l c \plmh}$) if quotient $p_\lph / \plmh = 10^w$ is constant; this is true for all bins except the cutoff ones. This allows us to use a preliminary solution mapping method for $\ql$, similar to that in Section \ref{sect:cutoff_momenta-approx}. At initialization a solution array is found. 
The value of $\ql$ obtained from the array can be used as is or further refined using 1D numerical scheme.

\section{Initial and boundary conditions}\label{sect:initial_condition}

\subsection{Initial spectrum}\label{sect:initial_spectrum}
\begin{figure}
   \includegraphics[width=1.\columnwidth,clip, trim=0.45cm 0cm 0.0cm 0.15cm]{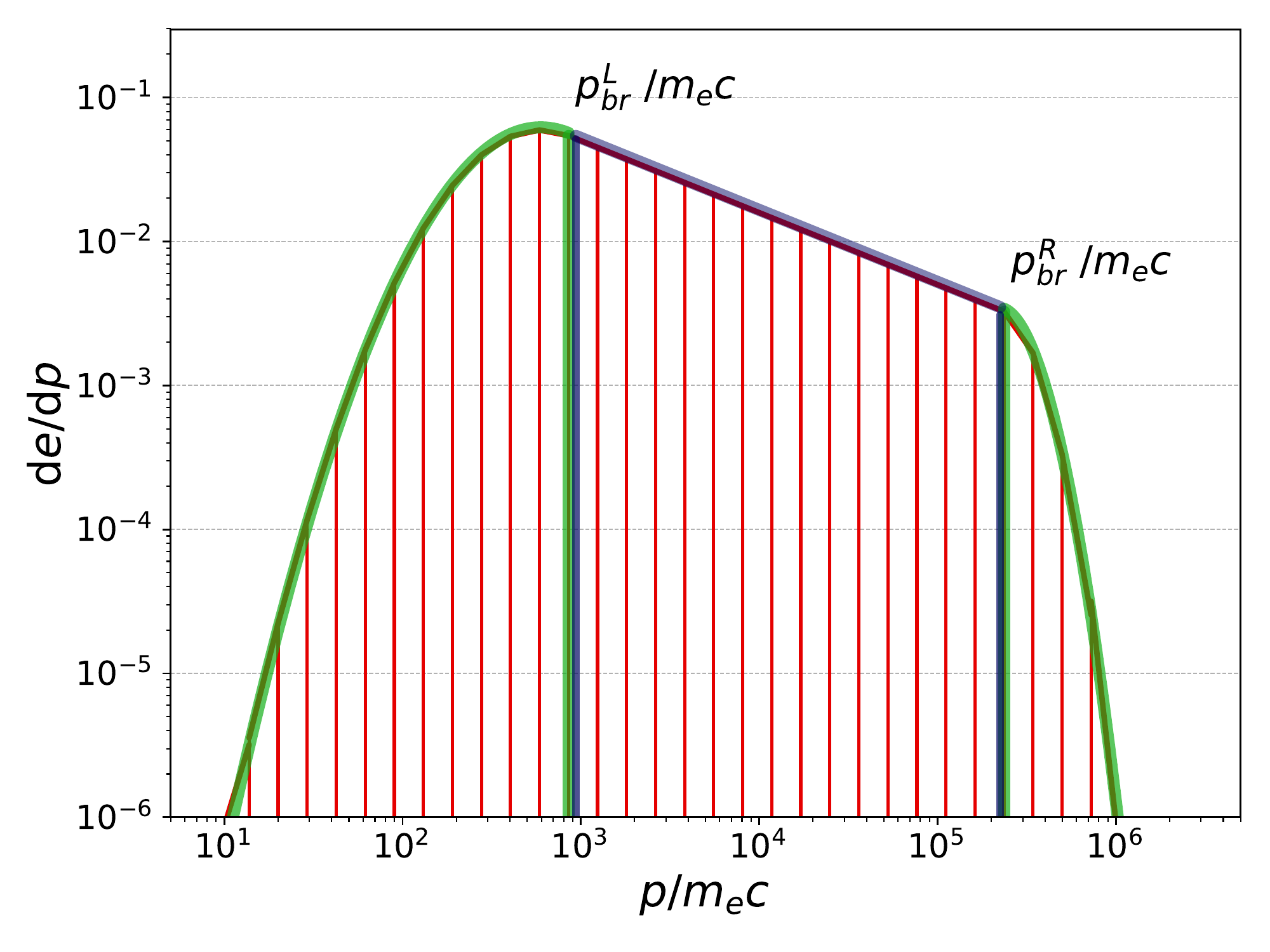}
   \caption{Example CR electron initial spectrum setup used in this paper, shown for differential energy density: analytical power law part (marked blue) and ''parabolic'' cutoffs (marked green) with its numerical implementation from CRESP (red).} \label{fig:e_initial_condition_plpc}
\end{figure}

The simplest condition for the power-law-type initial spectrum would be the pure power-law function. For convenience of implementation of the numerical algorithm, we shall use the momentum coordinate and related quantities of their dimensionless form:
\begin{equation}
   \pdl = p / (m_e c).
   \label{eq:p_dimensionless}
\end{equation}
We may consider a simple, purely power-law form of initial distribution function 
\begin{equation}
   f(\p) = \left\{
   \begin{array}{c c c}
      f_{\textrm{init}} \left(\frac{\p}{\p_{\textrm{lo}}}\right)^{-q\limr{init}}  & \mbox{for} & \p \in [\pdllo, \pdlup]\\
      0                   &               &    \mbox{elsewhere}\\
   \end{array}\right. .
   \label{eq:init_powerlaw}
\end{equation}
However, the initial spectra with a sharp cutoffs (as shown in Figure~\ref{fig:example_spectrum_e}) are impractical for numerical treatment. Realistic spectra, such as those  measured in the Solar System, also do not show abrupt changes in slope \citep[e.g.,][]{PhysRevLett.113.221102, Cummings_2016}. We assume a smooth source distribution function which:
\begin{enumerate}
   \item represents a power-law  in the range $(\pdlbrr;\pdlbrr)$ with slope $q\limr{init}$,

   \item is smooth near the two matching points $\pdlbrl$ and $\pdlbrr$,

   \item  falls  mildly, outside the range $(\pdlbrl;\pdlbrr)$, towards the lower and upper cutoffs $\pdllo$ and $\pdlup$, where it reaches its lowest values corresponding to $\esmall$.
\end{enumerate}
These requirements can be fulfilled by assuming that the two matching functions are second order polynomials in $\logba{\p}{10}$, while the power-law part of the spectrum is a first-order polynomial. Smooth matching of the three parts of the spectrum can be ensured by requiring continuity of the distribution function and its first derivatives at $\pdlbrl$ and $\pdlbrr$.
The appropriate distribution function can be defined with $\pdl$ as follows:
\begin{equation}
   f\limr{inj}(\p) = \left\{
   \begin{array}{l c l}
      f\limr{init} \left( \frac{\p}{\p\limr{lo}} \right)^{-q\limr{init}}                 & \; \mathrm{if} \; &  \p \in (\pdlbrl, \pdlbrr) \\
      \funit \; 10^{F(\logba{\p}{10})}   & \; \mathrm{if} \;  &  \p \in \left[ \pdllo,\pdlbrl \right]  \\
      & \mathrm{or} & \p \in\left[ \pdlbrr,\pdlup \right] \\
      0 &  &\mathrm{otherwise} \\
   \end{array}
   \right.
\end{equation}
where $\funit$ is constant with value of one and has the physical dimension of distribution function $f$ in the currently used unit system. 
For galactic models we use the system based on Solar mass $\Msun$, parsec $\pc$, and one million years $\Myr$, which implies 
$\funit = 1 \, \Myr^3 \Msun^{-3}\pc^{-6}$. 
The function
\begin{equation} 
 F(\logba{\p}{10}) = \alpha_{L,R} \left( \logba{\p}{10} \right)^2 + \zeta_{L,R} \logba{\p}{10} + \lambda_{L,R} 
\end{equation} 
depends on coefficients  $\alpha_{L,R}$, $\zeta_{L,R}$ , $\lambda_{L,R}$  which should be adopted to ensure smooth matching of the distribution function at each of $\pdlbrl$ and $\pdlbrr$ by solving the system of equations
\begin{eqnarray}
   \left\{
   \begin{array}{ l l}
      \logba{\left[\frac{f_\textup{init}}{\funit} \left( \frac{\p_\textup{br}^{\L,\R}}{\p_\textup{lo}} \right)^{-q_\textup{init}} \right]}{10} & =  F\left(\logba{\p_\textup{br}^{\L,\R} }{10} \right), \\[7pt]
      \logba{\Big[ \frac{\esmall}{4 \pi c (\p\limr{lo,up} \, m_e c)^{3} \funit}\Big]}{10} & =  F\left( \logba{\p_\textup{lo,up}}{10} \right), \\[7pt]
      \pder{}{\p}{} \logba{ \Big[ \frac{f_\textup{init}}{\funit} \left( \frac{\p}{\p_\textup{lo}} \right)^{-q_\textup{init}} \Big] }{10} _{\p\limr{br}^{\L,\R}} & =  \pder{}{\p}{} \left[ F\left( \logba{\p}{10} \right) \right]_{\p\limr{br}^{\L,\R}}.
   \end{array}
   \label{eq:initial_spectrum}
   \right.
\end{eqnarray}
where $f\limr{init},\, q\limr{init},\, \esmall,\, \pdllo,\, \pdlup,\, \pdlbrl$, and $\pdlbrr$ are parameters.
For convenience we define other dimensionless quantities:
\begin{equation}
   \fdl = f / \funit    \qquad \mathrm{and} \qquad   \edlsmall = \esmall / \epsunit, \nonumber
\end{equation}
where $\epsunit$ has the value of unity and dimension of spectral energy density of $\epsilon$ (e.g., $1 \,\Myr^{-1} \pc^{-2}$ in the currently used galactic unit system).
Setting an example with $\fdl\limr{init} = 1 $, $q\limr{init} =4.1$, $\edlsmall = 10^{-6}$, $\pdllo = 10$, $\pdlup = 10^6$, $\pdlbrl \approx 10^3 $, and $\pdlbrr \approx 10^5$ we obtain $\alpha_\L \approx -2.46$, $\zeta_\L \approx 10.34$, $\lambda_\L \approx -17.05$ for the left part of the spectrum $\alpha_\R = -6.07 $, $\zeta_\R \approx 57.18$, and $\lambda_\R \approx -150.54$ for the right part.
The example is illustrated in Figure~\ref{fig:e_initial_condition_plpc}.
\subsection{Injection of CR population in SN remnants}

The series of tests, described in Section \ref{sect:piernik_tests}, is initiated with a single SN explosion approximated by a local injection of CR protons centered at the point $(x_0,y_0,z_0)=(0,0,0)$ with an initial Gaussian profile of the CR proton energy density
\begin{equation}
   \begin{aligned}
      e_{\textrm{CRp}}(x,y,z) & = A_{\textrm{CRp}} \times
      \\ \times \; \mathrm{exp}  \bigg(-  & \frac{(x-x_0)^2+(y-y_0)^2 + (z-z_0)^2}{r_0^2} \bigg),
   \end{aligned} 
   \label{eq:initial_gaussian_spatial_distribution}
\end{equation}
with
\begin{equation}
   \begin{aligned}
      A_{\textrm{CRp}} &= \frac{\crpeff E\limr{SN}}{\pi^{3/2} r_0^3}  = \\
      = 9.4 &\times 10^6 \crpeff \left( \frac{E\limr{SN}}{10^{51}\erg}\right)\left( \frac{r_0}{1\,\pc} \right)^{-3} \hspace{-5pt} \Msun\, \pc^{-1} \Myr^{-2}
   \end{aligned}
   \label{eq:amp_cr_protons}
\end{equation}
where $r_0$ is the SN remnant radius determining the sphere where most of CRs are produced.
We assume that 10\% ($\crpeff = 0.1$) of the canonical SN kinetic energy output  $E_{\textrm{SN}}= 10^{51} \erg$ is converted to CRs. The normalization factor ensures that the spatial distribution integrated over  volume brings the total energy input of CR protons equal to $\crpeff E\limr{SN}$.
Initial total (bin integrated) CR electron energy density $e_{\mathrm{CRe}}(x,y,z)$ and the distribution function $f\limr{inj}(x,y,z,p)$ are scaled with respect to the energy density of CR protons
\begin{equation}
   \begin{aligned}
      e_{\mathrm{CRe}}(x,y,z) & = \creeff \times e_{\mathrm{CRp}}(x,y,z) \\
      &= \int_{p^\mathrm{(lo)}\limr{inj}}^{p^\mathrm{(up)}\limr{inj}} 4\pi p^2 T(p) f\limr{inj}(x,y,z,p) dp,
   \end{aligned}
   \label{eq:cresp_amplitude_scaling}
\end{equation}

where $e_{\mathrm{CRe}}$ and $\creeff$ are  energy density and production efficiency (relative to protons) of CR electrons.
We assume $\crpeff = 0.1$  and  $\creeff = 0.01$  \citep[see, e.g.,][]{2007ARNPS..57..285S}.
Having the distribution function amplitudes and slopes computed for every bin with the aid of formula~(\ref{eq:cresp_amplitude_scaling}), we determine
the initial particle number density $\nl$ and energy density $\el$ for each cell $(i,j,k)$.

\subsection{Boundary conditions for CRs on outer boundaries of the spatial grid} \label{sect:boundary_conditions}
Boundary conditions for the thermal fluid and magnetic field are realized in the standard way by copying the contents of respective cells to boundary cells. Periodic and internal boundary conditions for all the cosmic ray variables are realized in the same way as for the thermal fluid, i.e., by copying the contents of a number of grid zones  adjacent to the boundary (by default four layers of boundary cells on each side of the physical domain).
At outer boundaries, such as the outer $z$-boundaries in the wind test problem described in Section~\ref{sect:piernik_crwind_test},
we apply the diode-outflow boundary conditions for the thermal fluid, which  permit thermal plasma to flow out  of the computational domain, but not to flow in. In the case of CRs we apply boundary conditions typical for the diffusion problem, by setting the zero value of each CR variable (number densities and energy densities in each momentum bin) on each outer boundaries of the computational domain.
\section{Isolated CRESP tests} \label{sect:cresp_tests}
In this section we present elementary tests of isolated CRESP algorithm. We performed a series of runs, in which the CR electron population in a single cell was exclusively subjected to adiabatic expansion/compression or synchrotron losses. We aim to compare the results with analytical solutions for these elementary processes and additionally compare results obtained using the method described in Section \ref{sect:cutoff_momenta-approx}, to which we refer as \textit{approximated} approach (cutoffs are recovered at the beginning of each step) against the unapproximated cutoffs case (also referred to as \textit{exact} cutoffs), where the cutoffs are stored in the memory. No spatial transport 
is accounted for at this stage, and therefore no cutoff migration takes place, making the unapproximated cutoff case tests possible to carry out. We refer to the paper by \citet{1962SvA.....6..317K}, where an analysis of the effect of various modes of energy losses were studied using analytical methods. We compare numerically obtained spectra and spectrum cutoffs for synchrotron process with these analytical solutions.
\subsection{Initial spectrum}
Table \ref{tab:initial_params-isolated} presents all variable parameters of the initial spectrum for all tests in this section; parameters not listed  there 
remain unchanged in these experiments: $\nbin = 45$, $\p_{\mathrm{lo}} = 10 $, $\p_{\mathrm{up}} = 10^6$, and $\pdlbrr = 2.5 \times 10^5$.
\begin{table}[h]
   \small
   \centering
   \begin{tabular}{l l c c r r}
      \toprule
    & $\fdl\limr{init}$ & $q\limr{init}$&  $u\limr{d,0}$ &  $B_\perp$($\mu$G) &  $\p_{\mathrm{br}}^{\L}$ \\
   \toprule

   A1p         &    $1 \times10^{-9}$ & 3.5 &  -0.2 & 0.  & $2 \times 10^2$ \\

   A2p         &    $1 \times10^{-9}$ & 4.1    &  -0.2 & 0.  & $ 2 \times 10^2$ \\

   \bf{aA1p}   &    $1 \times10^{-9}$ & 3.5 &  -0.2 & 0.  & $ 2 \times 10^2$ \\
   \bf{aA2p}   &    $1 \times10^{-9}$ & 4.1   &  -0.2 & 0.  & $ 2 \times 10^2$ \\

   \hline

   S1          &    $5 \times10^{-11}$ & 3.5  &  0.   & 25.  & $10^3$ \\

   \bf{aS1}    &    $5 \times10^{-11}$ & 3.5  &  0.   & 25.  & $10^3$  \\

   S2          &    $7 \times10^{-12}$ & 4.4    &  0.   & 25.  & $10^3$ \\

   \bf{aS2}    &    $7 \times10^{-12}$ & 4.4    &  0.   & 25.  & $10^3$ \\

   \bottomrule
   \end{tabular}
   \caption{Numerical values of initial parameters chosen for various tests. All runs were conducted assuming $\nbin = 45$, $\p_{\mathrm{lo}}=10$, $\p_{\mathrm{up}} = 10^6$, $\pdlbrr = 2.5 \times 10^5$ and $\edlsmall = 10^{-6}$, $q$ denotes spectrum slope for the power-law part, $\pdlbrl$ and $\pdlbrr$ denote spectrum matching points corresponding to particle momenta. Adiabatic tests names start with A (''p'' for periodically variable $u\limr{d}$); synchrotron cooling tests with S. Names of runs with approximated cutoffs (bold) start with an additional ``a''.}\label{tab:initial_params-isolated}
\end{table}

\subsection{Adiabatic evolution test} \label{sect:adiab_iso_test}
Periodic adiabatic compression/expansion experiments aim to test algorithm's ability to preserve spectrum shape after multiple passages through bin faces and the impact of the approximation of cutoffs on the spectrum. 

We define time dependent velocity divergence:
$u\limr{d}(t) \equiv  1/3 \; \nabla \cdot \mathbf{v}   =  -u\limr{d,0}\cos (\omega\limr{d} t)$, which implies the following solution of equation (\ref{eq:btot}):
\begin{equation}
   p(t) = p\limr{t=0} \, \exp(-u\limr{d,0} \, \sin (\omega\limr{d} t) / \omega\limr{d}) \label{eq:time_dependent_ud}
\end{equation}
where the value of $\omega\limr{d}$ is set so that each adiabatic compression/expansion cycle takes $T=40$ in code units.
Adiabatic evolution should not change the overall spectrum shape, which should return to its initial shape at the end of each adiabatic cycle.
Periodic adiabatic compression/expansion tests,  named A- for exact and aA- for  approximated cutoffs, were conducted for two various slopes $q$ in the power-law part of the spectrum. 
\begin{figure}
   \includegraphics[width=\columnwidth, clip, trim=0.45cm 0.55cm 0.35cm 0cm]{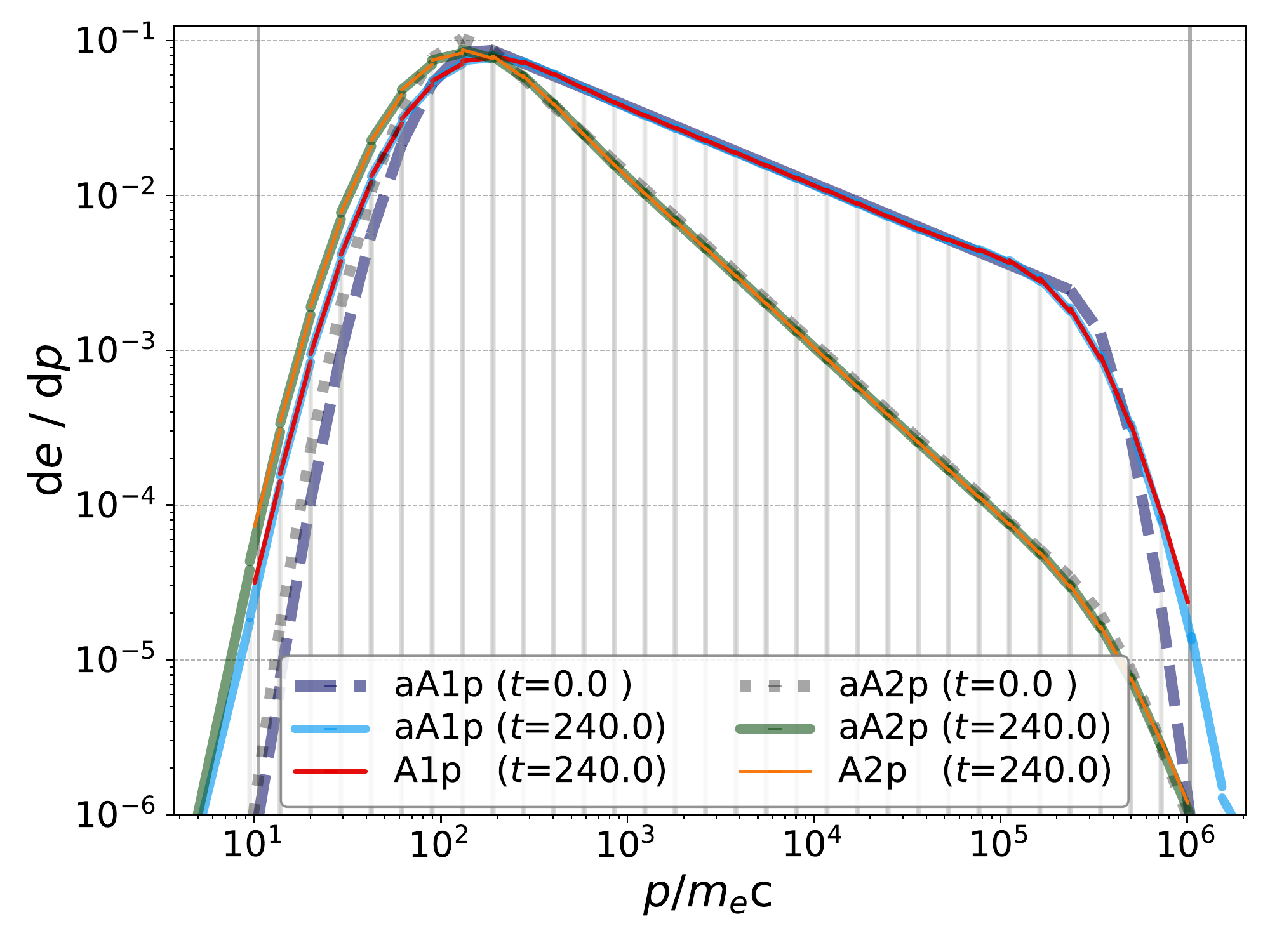}
   \caption{Depiction of spectral energy density in tests A1p/aA1p and A2p/aA2p, comparing the spectra for $t = 0$  (dark-blue/grey, dashed/dotted; tests pairs A- and aA- start with the same initial spectrum) and the final stage (solid, red/orange for exact and blue/green for approximated cutoff case) after six adiabatic compression/expansion cycles, which corresponds to $t=240$ in code units.).
   Vertical gray lines mark the initial cutoffs (equation \ref{eq:p_t+dt}).}
   \label{fig:e_adiabatic_evolution}
\end{figure}

In Figure~\ref{fig:e_adiabatic_evolution} we present spectral distribution of particles energy density from tests A1p, aA1p, aA2p, and A2p at initialization and after $t=240$ in code units. In the range described by the pure power-law ($\p_{\mathrm{br}}^{\L}$ --- $\p_{\mathrm{br}}^{\R}$), all spectra remain almost unaffected and preserve the initial slope throughout all six cycles. Beyond the pure power-law range spectra slightly diffuse with time in all tests, which is more distinct at momenta lower than $\p_{\mathrm{br}}^{\L}$ and noticeable beyond $\p_{\mathrm{br}}^{\R}$ in aA1p test. The low momentum cutoff $\p_{\mathrm{lo}}$ shows a falling trend in tests with cutoff approximation; in aA1p test its value decreases by $22.1 \% $ ($\sim3.7\%$ per cycle), while in aA2p the decrease yields $\sim 21.8 \%$ test ($\sim 3.6\% $ per cycle). The high momentum cutoff $\p_{\mathrm{up}}$ in aA1p reveals an increase in time by $\sim 6.1\%$ per cycle on average ($\sim 36.4\%$ total), while in aA2p it shows only slight variations ($<1\%$ per cycle). In tests without approximation the numerical cutoffs match analytical ones almost perfectly. However we note that a slow diffusion of the spectrum is also present here.

The change in cutoff momenta is clearly a consequence of the used approximation method, described in Section~\ref{sect:cutoff_momenta-approx}, which implies a  dependence of cutoffs on the height of the spectrum above the base level $\esmall$  as well as on the spectrum shape.
We find that the  algorithm preserves very well the total CR energy density, while our approximation of cutoff evolution only moderately influences the cutoff positions in the case of periodic adiabatic compression-expansion process.
\subsection{Synchrotron cooling test}
\begin{figure}
   \hspace{-15pt}\belowbaseline[-3.15pt]{\begin{tikzpicture}[spy using outlines={rectangle, magnification=2.3,connect spies}]
      \node {\includegraphics[height=6.4cm, clip, trim=0.2cm 0.25cm 0.25cm 0cm]{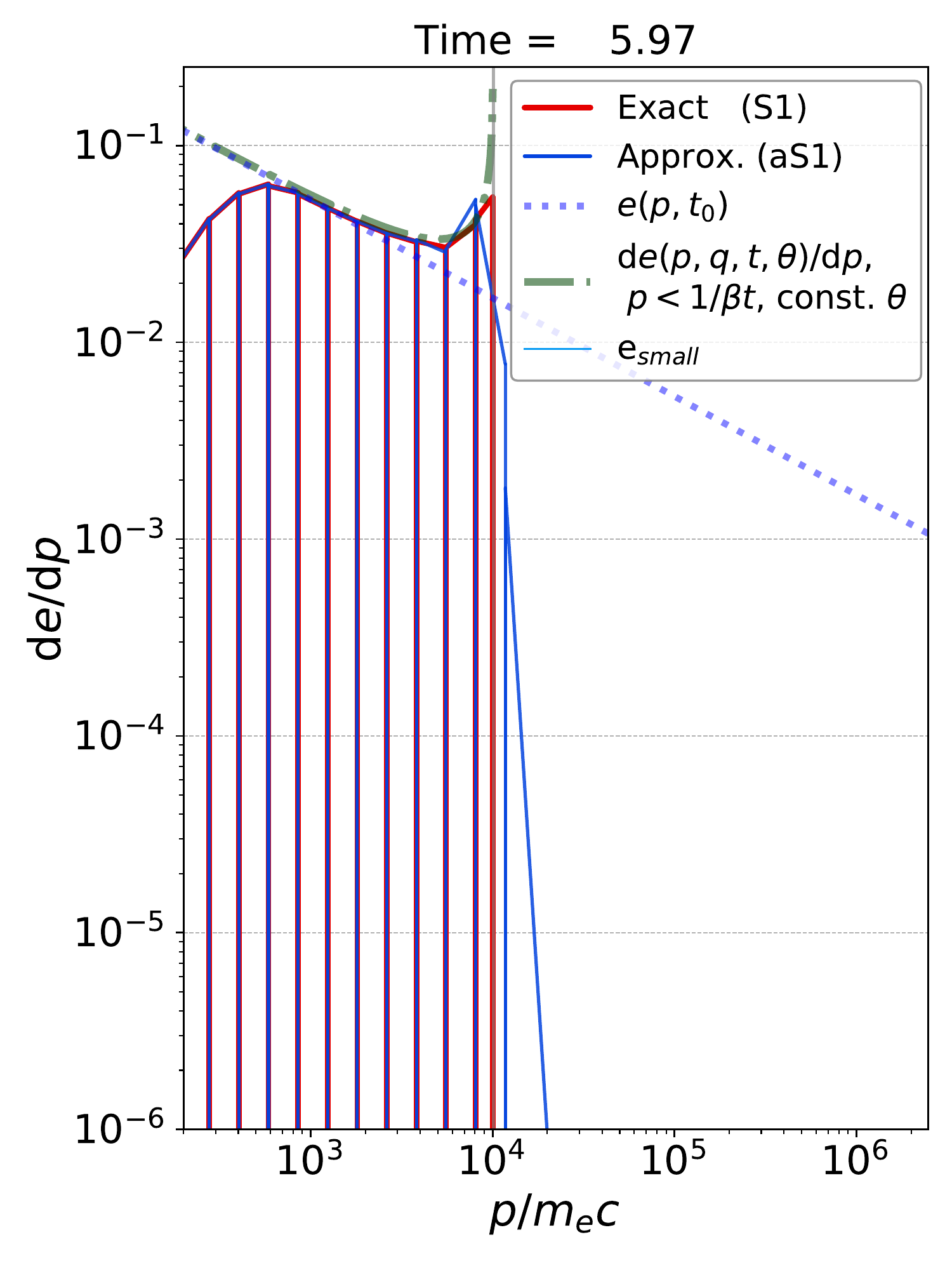}};
      \path (-0.155, 1.9) coordinate (spypoint)
            (1.4,-1.15) coordinate (spyviewer);
      \spy[gray, width=1.5cm,height=2.5cm] on (spypoint) in node [fill=white] at (spyviewer);
   \end{tikzpicture}}\hspace{-12pt}
   \belowbaseline[-3.15pt]{\begin{tikzpicture}[spy using outlines={rectangle, magnification=2.3,connect spies}]
      \node {\includegraphics[height=6.4cm, clip, trim=2.75cm 0.25cm 0.25cm 0.cm]{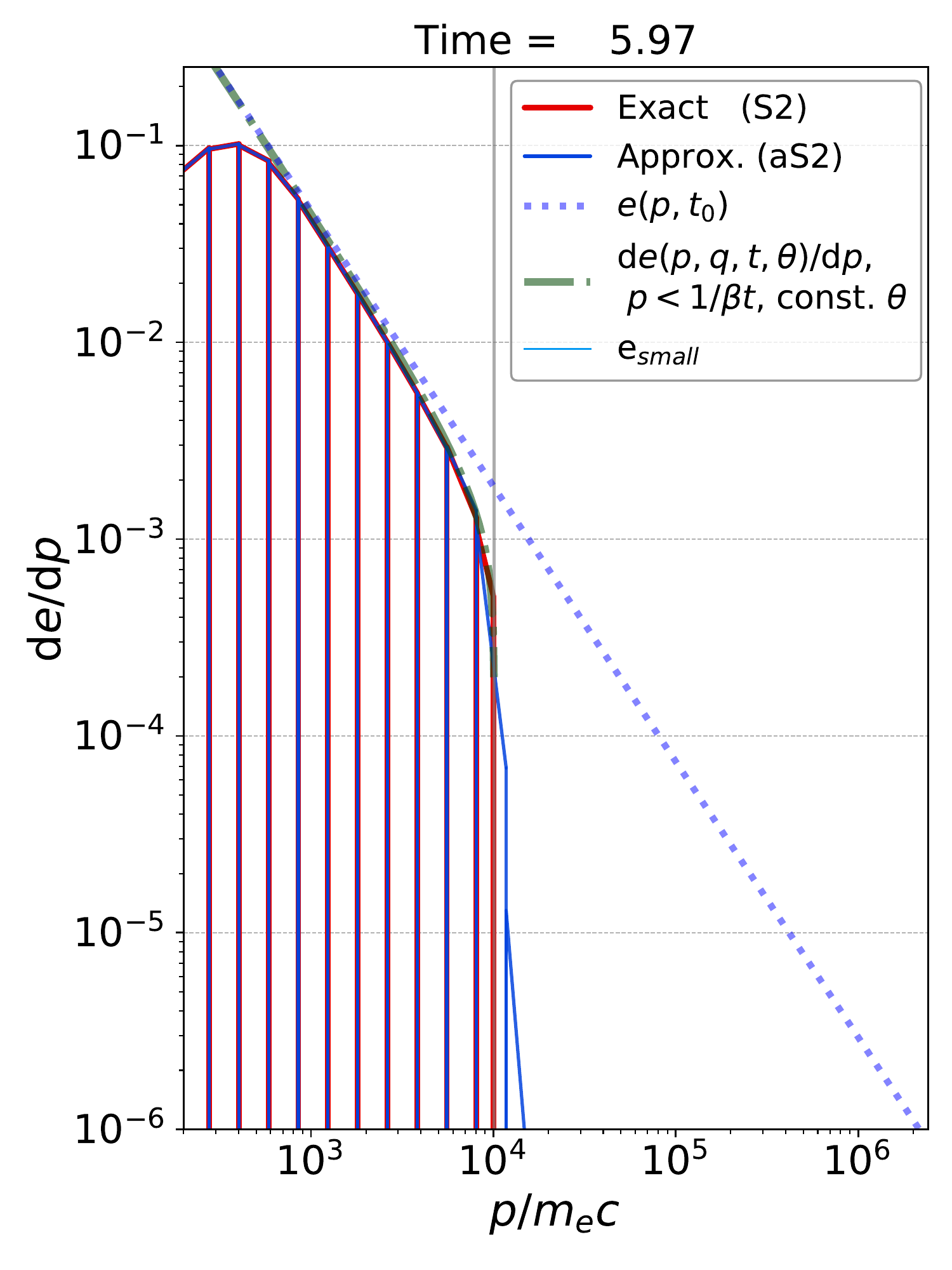}};
      \path (-0.455, 0.615) coordinate (spypoint1)
            (1.05,-1.15) coordinate (spyviewer1);
      \spy[gray, width=1.5cm,height=2.5cm] on (spypoint1) in node [fill=white] at (spyviewer1);
   \end{tikzpicture} }
   \vspace{-5pt}\caption{Synchrotron losses of CR electron population in energy density, with relevant momentum range shown. Spectra with $q = 3.5$ (S1--aS1, left) and $q = 4.4$ (S2--aS2, right), $\edlsmall=10^{-6}$, dotted lines mark the initial slope, and both the exact (red) and approximated (blue) cutoff cases are shown. The vertical gray line marks synchrotron energy limit $1 / \beta t$ (as in equation (\ref{eq:cr_transp_df_anisotropic_t})), $\esmall$ marks the bottom level.
   The analytical curves for synchrotron cooling (green, dot-dashed) were taken from \citet[equation 5]{1962SvA.....6..317K}. }
   \label{fig:synch_cooling_e_comparison}
\end{figure}

Considering particles with the same pitch-angle $\theta$, we take the formula from \citet[equation 5]{1962SvA.....6..317K} that describes the evolution of the power-law energy spectrum of CR electrons distribution $N(E,t)$ subject to synchrotron losses. This solution  expressed in terms of distribution function reads:
\begin{equation}
   f(p,q,\theta,t) = \left\{
   \begin{array}{l l}
      f(p)(1 - \beta t p)^{q - 4},  & p < \frac{1}{\beta t} \\
      0, & p > \frac{1}{\beta t}
   \end{array}
   \label{eq:cr_transp_df_anisotropic_t}
   \right.,
\end{equation}
where $\beta = \frac{2 \sigma_T}{m_e^2 c^2} \frac{2 \, \left(\sin\theta \; B \right)^2} {8 \pi} $.
The formula implies that for $q < 4$ the spectrum rises with time and falls abruptly to zero around $p \sim 1 / \beta t$, while for $q > 4$ downward steepening of the spectrum in high-momentum range should occur.
Within the current setup of isolated-cell (with no spatial transport of CR particles), we performed a series of synchrotron cooling tests, aiming to examine the algorithms' ability to reproduce expected features and cutoff evolution. In our later tests, the distribution of particle momenta is assumed to be isotropic, hence pitch-angle integrated synchrotron cooling coefficient $\ubr$ from equation (\ref{eq:btot}) is used; however, here for simplicity we assume $\theta = \pi / 2$. These experiments, named S- (exact) and aS- (approximated cutoffs), were conducted for $q < 4$ (S1, aS1) and $q > 4$ (S2, aS2).
\\We compare the spectra in Figure~\ref{fig:synch_cooling_e_comparison} at $t=$ 5.97 Myr, being the time needed for $\p_{\mathrm{up}} = 10^6$ of CR electrons to cool by factor of 100 in the presence of $B_\perp = $ 25 $\mu$G (equation \ref{eq:p_t+dt}). The analytical curve of energy density is obtained using equations~(\ref{eq:e_l}) and (\ref{eq:cr_transp_df_anisotropic_t}), plotted with green dash-dotted line on the figures.
In both cases the evolution of high-energy end of the spectra reproduces the analytical results
of \citet{1962SvA.....6..317K}.
We expected $\p_{\mathrm{up}} = 10^4$ and found the numerical values to be $\p_{\mathrm{up,S1}} = 1.05\times 10^4$, $\p_{\mathrm{up,aS1}} = 1.85\times 10^4$, and $\p_{\mathrm{up,aS2}} = 1.53\times 10^4$ respectively. The departures of the approximated solutions with respect to the exact one are comparable to the bin width, therefore we consider the approximation error to be satisfactory small.
The growth of departures from analytical cutoff in aS- tests with time is rather slow, indicating that synchrotron cooling naturally limits the numerical diffusivity of the cutoff approximation algorithm.

\section{Tests of CRESP with PIERNIK}\label{sect:piernik_tests}
In this section we present tests of the CRESP algorithm embedded in the PIERNIK MHD code.
The fluid transport algorithms of PIERNIK are used in the advection and diffusion steps of CRs on the spatial Eulerian grid, while the CRESP algorithm is used in each spatial cell to transport CRs on the momentum grid.
We prepared a series of tests aiming to check the interoperation of CRESP algorithms with transport of the entire spectrum on the spatial grid by the fluid transport algorithms of PIERNIK. We shall switch on and off individual processes: advection, diffusion, adiabatic process, and synchrotron losses.
The present setup assumes instantaneous injections of CR protons and electrons in isolated supernova remnants.

{ CR protons play only  an auxiliary role in the presented tests, as drivers of thermal gas dynamics, and are subject to anisotropic diffusion, advection, and adiabatic process, modeled with the aid of momentum integrated diffusion-advection equation~(\ref{eq:e_all}),  typical for the two-fluid approach, using the algorithms described in \citep{2003A&A...412..331H, piernik4}.  Spectral evolution of the  CR proton component, omitted in this paper, will be addressed in a subsequent paper of this series.}

CR electrons evolve according to a momentum-dependent diffusion equation, integrated by means of the piece-wise power-law method by \citet{2001CoPhC.141...17M}, with the code extension, presented in this paper, designed for time-dependent transport on the $(x,y,z,p)$-grid of CR particles subject to the adiabatic process and synchrotron cooling mechanism.\\
In the advection and diffusion tests the domain size was $1\times1\times0.0625\; \kpc$ with spatial resolution $128\times128\times8$ grid cells and with periodic boundary conditions in all directions. For the adiabatic expansion test we assumed a domain with size 1$\;\kpc^3$ with 128$^3$ cells and outflow boundary conditions.
We set the initial spectrum of CRs extending over five decades in the momentum space ($\p $ in range $5 - 8.5 \times10^5$), with total 23 bins. The initial spectrum had amplitude $\fdl\limr{init} = 10^{-9}$ 
, slope $q=4.1$, and matching points at $\p_{\mathrm{br}}^{\L}=100$ and $\p_{\mathrm{br}}^{\R}=5 \times 10^5$, beyond which the spectrum was defined by means of second order polynomials in $\logten(p)$ -- see Section \ref{sect:initial_spectrum}. Spectrum cutoffs were approximated using the method described in Section \ref{sect:cutoff_momenta-approx}, with $\edlsmall = 10^{-10}$. 
CRs were injected by a single SN explosion with a radial profile given by equation~(\ref{eq:initial_gaussian_spatial_distribution}).
\subsection{Advection test}

\begin{figure}
   \belowbaseline[0pt]{\hspace{-12.5pt}\includegraphics[height=4.95cm,clip, trim=2.cm 0cm 0cm 0cm]{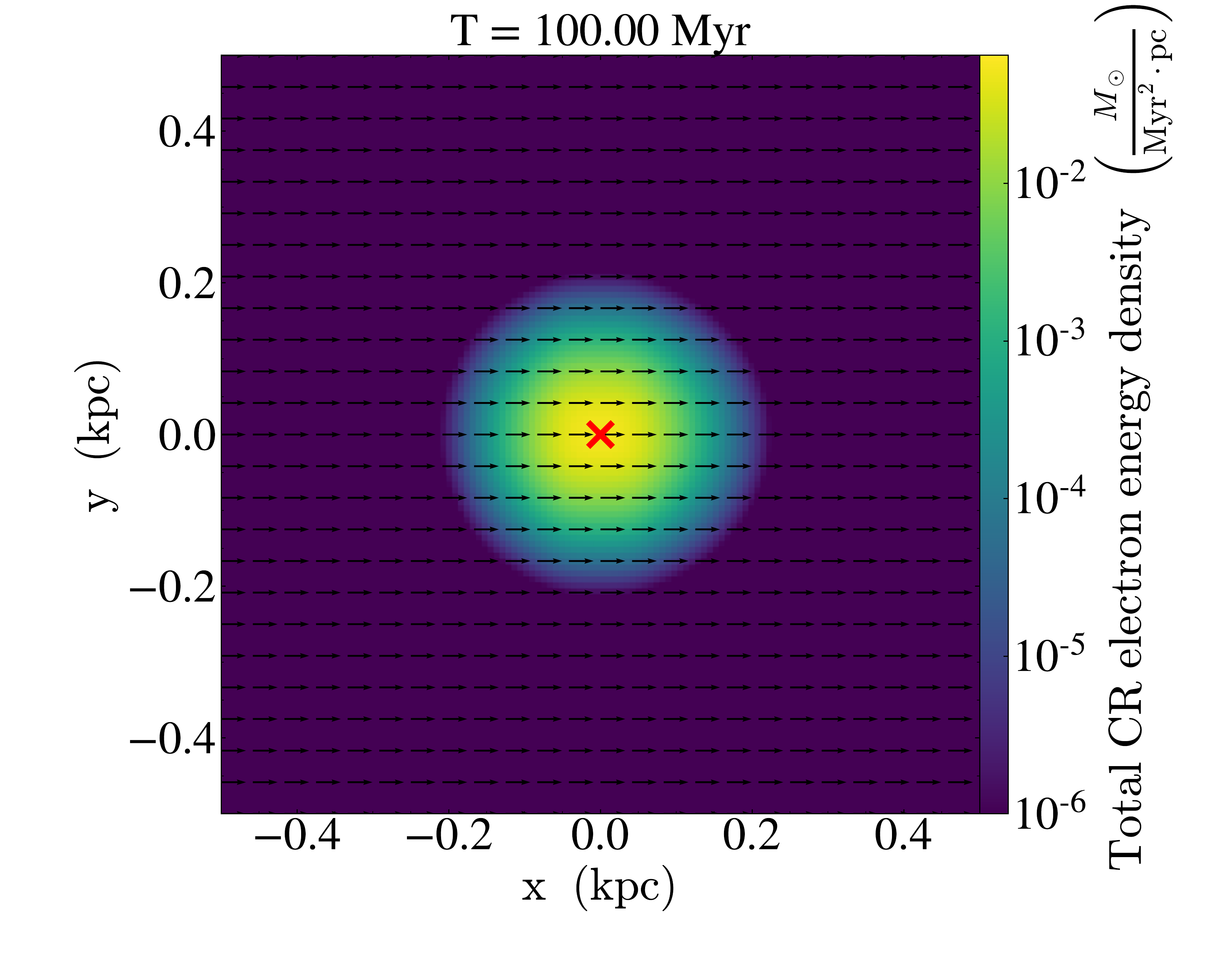}}\belowbaseline[3.5pt]{\hspace{-4pt}\includegraphics[height=4.5cm,clip, trim=0.1cm 0cm 0cm 0cm]{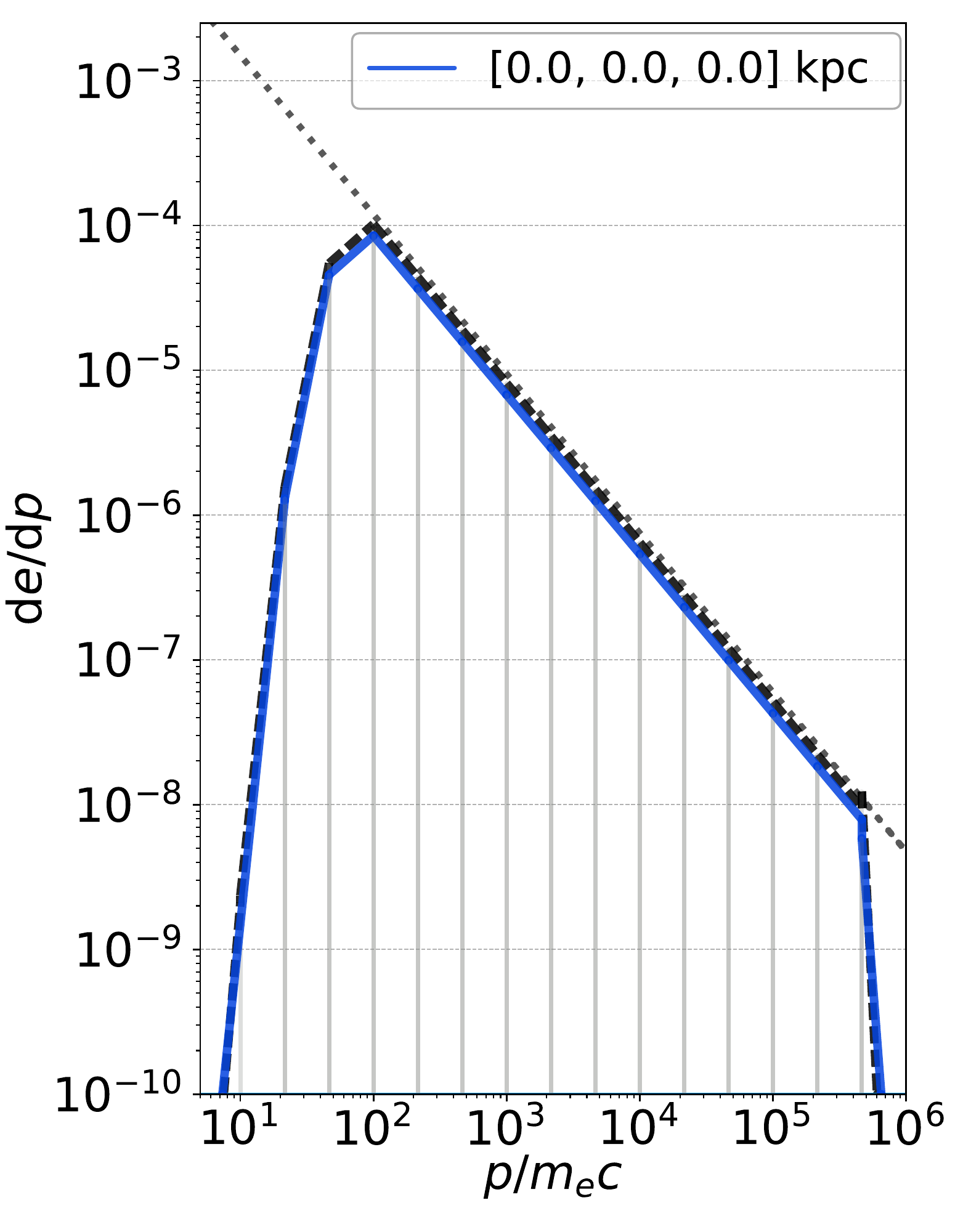}}
   \vspace{-10pt}\caption{Advection test results for $t = 100$ Myr, with velocity field vectors annotated. After 100 Myr of evolution the resulting spectrum is almost indistinguishable from the initial one, represented by the dashed black line onto which the resulting spectrum is superimposed.}
   \label{fig:piernik_advection}
\end{figure}

To test the pure advection of the whole energy spectrum in space, we set up velocity field $v_x = 10 \, \mathrm{ pc\; Myr}^{-1}, v_z = v_y = 0 $ in a periodic simulation domain and with diffusive transport and synchrotron losses switched off. The velocity field was uniform ($\nabla \cdotp \mathbf{v} = 0 $); no spectrum shape change was expected. 
The spatial and spectral distribution of energy density are shown in Figure~\ref{fig:piernik_advection} for $t = 100 \,\Myr$: the spectrum shape remains practically unchanged during the simulation time, except for a small reduction of the distribution function amplitude due to a relatively weak numerical diffusion. Comparing $n\limr{tot}$ and $e\limr{tot} $ for times 0 and 100 $\Myr$, we get relative errors of $\approx$ 0.17 for both quantities.

\subsection{Adiabatic expansion test}

\begin{figure}
   \belowbaseline[0pt]{\hspace{-20pt}\includegraphics[height=5.0cm]{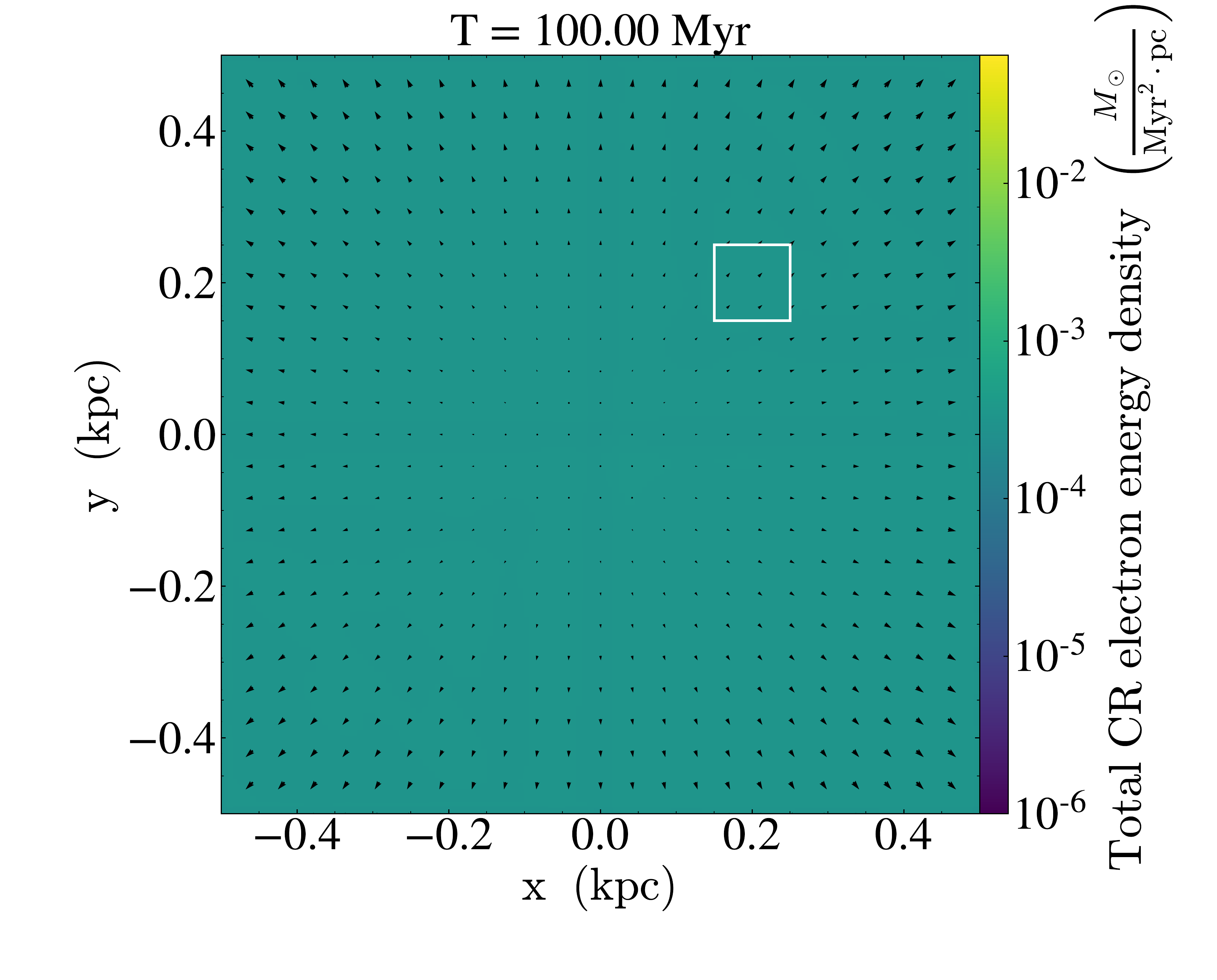}}\belowbaseline[5.5pt]{\hspace{-6pt}\includegraphics[height=4.55cm,clip,trim=0.1cm 0cm 0cm 0cm]{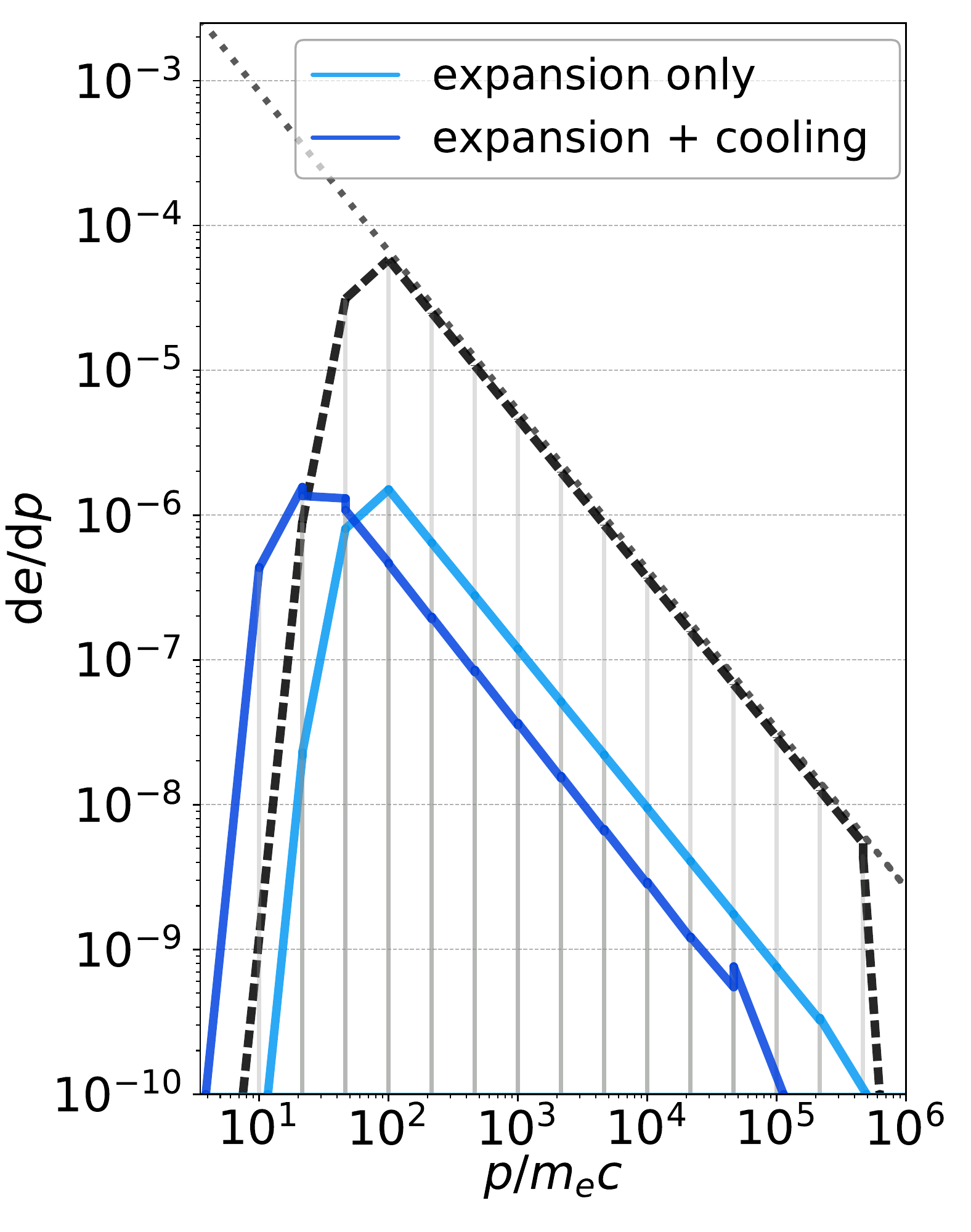}}
   \caption{Distribution of the total CR energy density in the adiabatic expansion test (with velocity field vectors annotated) and its spectra, averaged for the marked region (centered at [200, 200, 0] pc) at $t=0$ (black, dashed) and $t=100\, \Myr$: accounting only for the isotropic outflow -- expansion only (light blue, solid) and with adiabatic shift (blue, solid).}
   \label{fig:piernik_expansion}
\end{figure}

The adiabatic expansion test aims to validate the CR spectrum evolution under the action of adiabatic expansion of the underlying plasma and in the absence of diffusion and synchrotron losses. We assumed that spatial distribution of CRs in the domain in this test was isotropic, and initial velocity field of thermal gas was given by $\mathbf{v}(\mathbf{r}) = C \; (\mathbf{r}-\mathbf{r_c}) $, where $C = 0.023$ was the initial expansion rate which remained constant throughout the whole test. The distribution of $n$ and $e$ was uniform by setting high SN radius ($\sim 10^7 \, \pc$), implying computational domain coverage of only a small, central part of the SN remnant.
Distribution of CR energy density and the corresponding CR energy spectrum at $t=100$ Myr are presented in Figure ~\ref{fig:piernik_expansion}.\\
For a quantitative comparison of the numerical results, we take analytical solutions for momentum-integrated number and energy density diffusion-advection equations (\ref{eq:n_all}, \ref{eq:e_all}), in which all aforementioned assumptions are imposed for CR electrons. With assumed velocity field and $u_d \equiv \frac{1}{3}(\nabla \cdot \mathbf{v})$, these solutions yield:
\begin{eqnarray}
   n\limr{tot}(t+\Delta t) &  = & n\limr{tot}^{t=0} \; \mathrm{exp} \left(-3 \int_t ^{t+\Delta t} u_d(t) dt \right)  \nonumber \\
   e\limr{tot}(t+\Delta t) &  = & e\limr{tot}^{t=0} \; \mathrm{exp} \left(-4 \int_t^{t+\Delta t} u_d(t) dt \right) \nonumber
\end{eqnarray}
If $u_d$ remains constant in time, the latter simplifies to:
\begin{eqnarray}
   n\limr{tot}(t + \Delta t) &  = & n\limr{tot}^{t=0} \; \mathrm{exp} \left(-3 u_d \Delta t \right)  \nonumber \\
   e\limr{tot}(t + \Delta t) &  = & e\limr{tot}^{t=0} \; \mathrm{exp} \left(-4 u_d \Delta t \right) \nonumber
\end{eqnarray}
Quantities $n$ and $e$ stored in bins are already momentum integrated. Summing them over the active bins and comparison with analytical results for $t = 0$ and $100 \, \Myr$ yields relative errors of $\approx 0.12$ for $n\limr{tot}$ and $\approx 0.11$ for $e\limr{tot}$, comparable to relative errors in advection test.
These small deviations result plausibly from the fact that our present time-integration algorithm is only of the first order.

\subsection{Diffusion test} \label{sect:piernik_diffusion}
\begin{figure}
   \centerline{
   \belowbaseline[-12pt]{\hspace{-5pt}\includegraphics[height=5cm,trim={1.cm, 0 0 0 },clip]{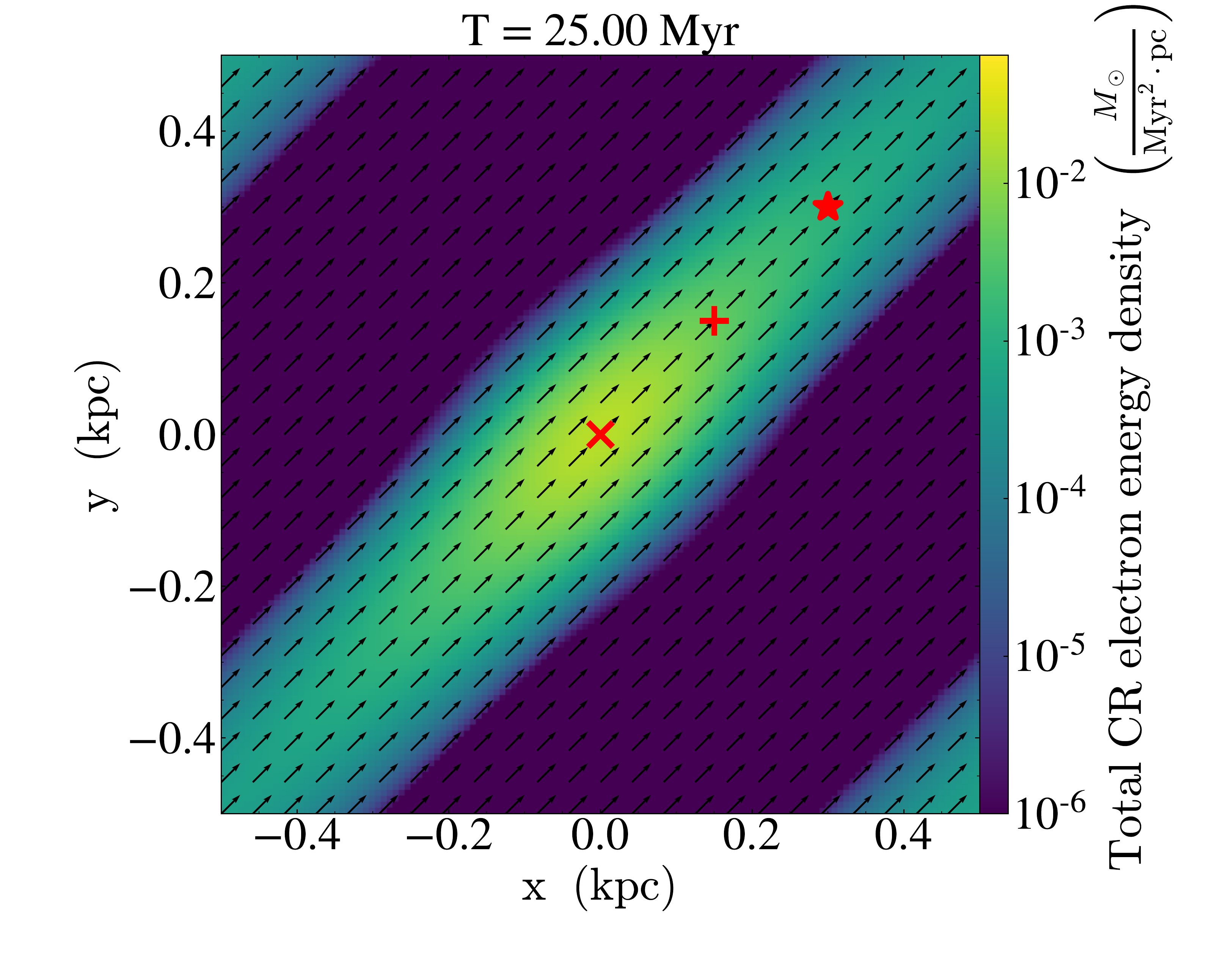}}
   \belowbaseline[-7.5pt]{\hspace{-8pt}\includegraphics[height=4.65cm, clip, trim=0.1cm 0cm 0cm 0cm]{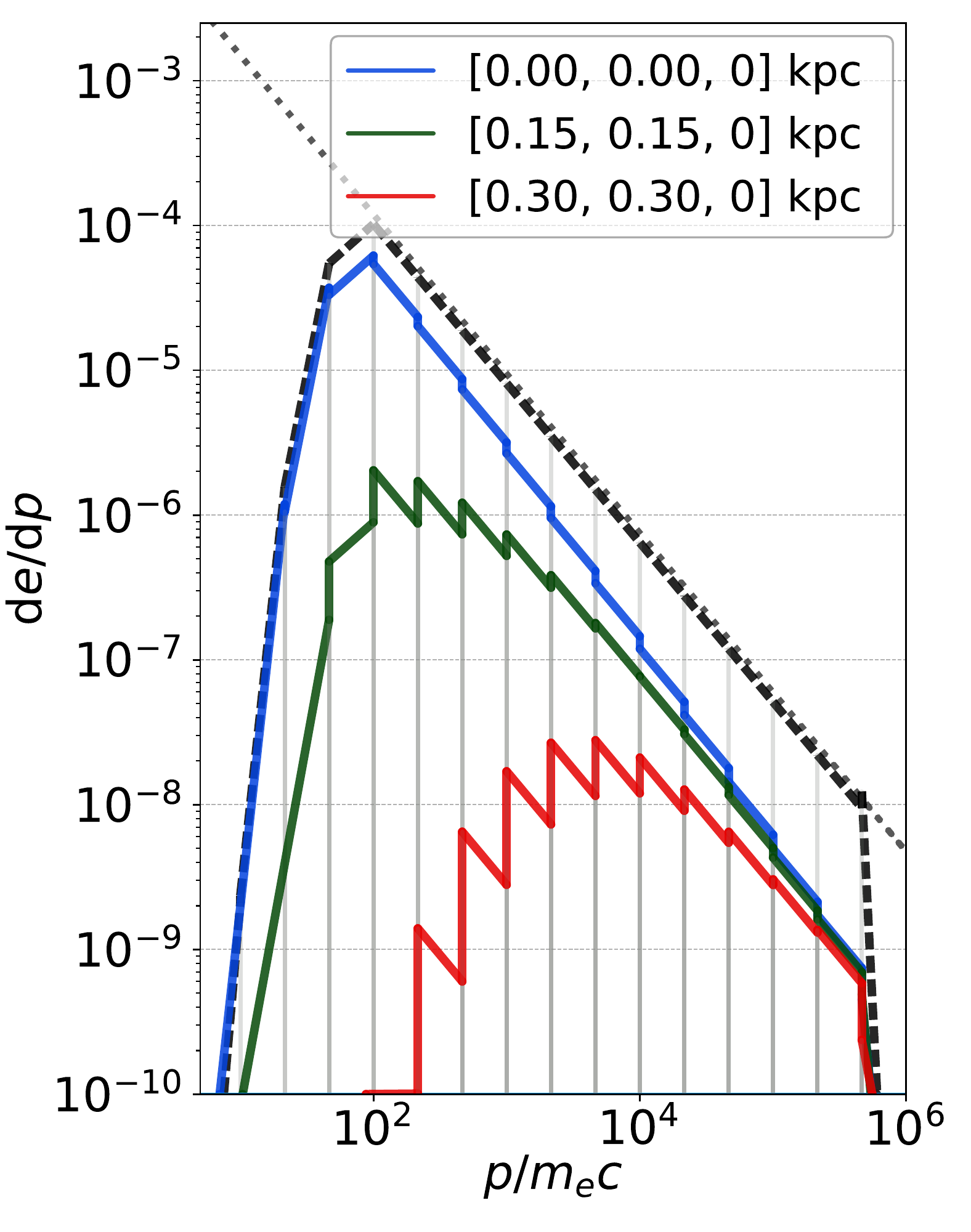}} }

   \caption{Energy dependent diffusion at  $t=25\; \Myr$: spatial distribution of total energy density with magnetic field lines annotated (left) and energy density spectra at marked positions (right). 
   The dashed black line in the right panel represents the initial spectrum at domain center; the blue, green, and red lines represent  spectra taken at $t= 25\;\Myr$ in the domain center  ('$\times$' mark),  $'+'$ and  $'*'$ points, respectively.
   The presented spectra show a softening effect near the injection point and a gradual hardening  growing with the distance from the injection center.
   The saw-shaped spectra at two off-center locations result from the simplified treatment of the diffusion process.}
   \label{fig:piernik_diffusion}
\end{figure}

The aim of the diffusion test was to validate the momentum-dependent diffusion part of CR spatial propagation algorithm in the absence of any other propagation effects. Diffusion coefficients were computed with formula (\ref{eq:diffusion-coefficients-approximate}), where $\kappa_{\parallel (10\mathrm{k})} = 3\cdot 10^{26} \, \cm^2 \, \s^{-1} \simeq 10^3 \, \pc^2 \, \Myr^{-1}$. $\kperp(p_l) = 10^{-3} \kparal (p_l) $ with momentum dependence given by $\alpha_\kappa = \frac{1}{2}$. The magnetic field was set to $b_x = b_y = 1 \, \muG$ (vectors shown on Figure~\ref{fig:piernik_diffusion}), synchrotron cooling was switched off for this test, and
the initial distribution of CRs in space was given by equation~(\ref{eq:initial_gaussian_spatial_distribution}). Momentum-dependent diffusion with coefficients defined in equation (\ref{eq:diffusion-coefficients-approximate}), treated separately from other propagation effects, causes the particles occupying separate bins to diffuse at different rates.
As the diffusion coefficients are approximated by their bin-centered values, the diffusion taken alone retains the slopes $q_l$ which are determined for a given ratio $e_\l/n_\l$ from equation~(\ref{eq:enpc-q}). This leads to the emergence of the saw-tooth pattern of the distribution function apparent in the right panel of Figure~\ref{fig:piernik_diffusion}.
Spectral indices of individual bins remain roughly unchanged, except near the cutoffs and lower momentum matching point.
A more detailed discussion of this point can be found in \citet{2020MNRAS.491..993G}.
\\The implication of the momentum dependent diffusion is the apparent softening of the spectrum near the injection point and hardening at larger distances. 
Away from the center ($\textcolor{black}{+}$ and $\textcolor{black}{\star}$ points) the spectrum  hardens, as expected, due to the growth of diffusion coefficients with momentum. At point $\textcolor{black}{\star}$ the spectrum is distinguishably dominated by the particles injected at the upper part of the momentum range.
Though the slopes $q$ are preserved in the bins, we observe hardening of the overall spectrum relative to the initial one with increasing distance from the center, while in the source center that spectrum becomes softer than the initial spectrum.

\section{CR-electron spectrum evolution in galactic wind}\label{sect:piernik_crwind_test}

In this section we analyze evolution of CR electron spectrum in a galactic wind driven by CRs.
We present the results of 3D simulations of an outflow driven by CR protons, supplemented with spectrally resolved CR electrons. This experiment demonstrates CR spectral evolution in a stratified box.
We assumed that:
\begin{enumerate}
   \item the CR-proton component, described by momentum-integrated diffusion-advection equation~(\ref{eq:e_all}), is  dynamically coupled with thermal gas and magnetic fields stratified by a vertical gravitational field of a galactic disk using the method described by \cite{2003A&A...412..331H}.

   \item the CR-electron component evolves according to the momentum-dependent diffusion-advection equations~(\ref{eq:nLR}) and (\ref{eq:eLR}) using the numerical algorithm developed in this paper based on the piece-wise power-law two-moment method \citep{2001CoPhC.141...17M}.

   \item the simulation volume represents a stratified box with domain sizes $(L_x,L_y, L_z) = (0.5 \, \kpc, 1 \,\kpc, 9 \,\kpc)$ and grid resolution $(n_x, n_y, n_z) = (24\times48\times432)$, with periodic boundary conditions used in horizontal directions and outflow-diode boundary conditions (see Section \ref{sect:boundary_conditions}) at the lower and upper computational box boundaries.

\end{enumerate}
Initial parameters for the ISM reflected galactic disk properties, including a vertical gravitational field model at galactocentric radius of 5$\,\kpc$, as taken from \citet{1998ApJ...497..759F, 2009A&A...498..335H}, yielding SN rate of 130$\,\kpc^{-2} \,\Myr^{-1}$, speed of sound 7 km s$^{-1}$, density of the ISM in the disk plane $\rho_0 = 0.125 \, \Msun \,\pc^{-3}$ (equivalent to hydrogen number density of $ \sim 5 \, \cm^{-3}$). We solved the hydrostatic equilibrium equation for the external gravitational field to get vertical distributions of gas and horizontal magnetic field, yielding $B_y \approx 1 \,\mu$G at $z=0$.
For electrons we set diffusion coefficients according to formula (\ref{eq:diffusion-coefficients-approximate}), assuming
${\kappa\limr{10\mathrm{k}}}_{\parallel}  = 3\cdot 10^{27}\cm^2 \, \s^{-1} \simeq 10^4$ pc$^2\,$Myr$^{-1}$ and $\kperp = 0.1 \kparal$  with momentum dependence power index $\alpha_\kappa = 0.0$ and $0.5$, with constant value of $\kparal = 10^4 \,\pc^2\,\Myr^{-1}$ and $\kperp = 0.1 \kparal$ for CR protons.

No cosmic rays were assumed in the initial state. SNe injected CR protons and electrons to the ISM with efficiencies $ \crpeff = 0.1$  and $\creeff = 0.01$ (defined in equation~(\ref{eq:initial_gaussian_spatial_distribution}--\ref{eq:cresp_amplitude_scaling})), referring to the
canonical kinetic energy output of a singe SN equal to $ E_{\mathrm{SN}} =  10^{51} \,\erg$, with radius $r_0 = 50 \,$pc.
We used the source spectrum of CR electrons defined by formula (\ref{eq:initial_spectrum}), assuming the following set of parameters:
$\fdl\limr{init}=10^{-2} $, 
$q\limr{init}=4.1$, $\p_{\mathrm{lo}}=5$, $\p_{\mathrm{up}}=8.5\times 10^5$, and matching points at $\p_{\mathrm{br}}^{\L} = 100 $ and $\p_{\mathrm{br}}^{\R} = 5\times10^5 $ with $\edlsmall = 10^{-8}$. 
To analyze the effects of separate processes, we ran a series of tests. The aim was to compare separately the effects of advection, diffusion, adiabatic, and synchrotron processes on the spectra and to compare the effects of energy-dependent and energy-independent diffusion. Selected processes, alternating parameters, and test names for CR-driven galactic wind simulation are presented in Table \ref{tab:mcrwind-test_names}.
\begin{table}
   \begin{tabular}{c c c c c c} \toprule
      \small
      \multirow{2}{*}{\parbox[t]{2mm}{\rotatebox[origin=l]{90}{Adv.}}} & \multirow{2}{*}{\parbox[t]{2mm}{\rotatebox[origin=l]{90}{Diff.}}} & \multirow{2}{*}{\parbox[t]{2mm}{\rotatebox[origin=l]{90}{Synch.}} } & \multirow{2}{*}{\parbox[t]{2mm}{\rotatebox[origin=l]{90}{Adiab.}}} & \multicolumn{2}{c}{Name} \\ \cline{5-6}
      & & & &  $\alpha=0.0$ & $\alpha=0.5$ \\
      \vspace{-11pt} \\
      \toprule
      + & - & - & - & \multicolumn{2}{c}{A} \\
      + & + & + & - & B0.0 & B0.5 \\
      + & + & - & + & C0.0 & C0.5 \\
      + & + & + & + & D0.0 & D0.5 \\
      \bottomrule
   \end{tabular}
   \caption{Collection of tests including various transport modes (Adv. -- advective, Diff. -- diffusive) and spectrum evolution processes (Synch. -- synchrotron, Adiab. -- adiabatic).}
   \label{tab:mcrwind-test_names}
\end{table}
\\Advective transport is essential to the wind process and was therefore included in every test.
In test A it was the only considered process; we expected the spectral slopes $q$ to remain unchanged.
In test B we included advective and diffusive transport along with synchrotron process. We expected spectrum to manifest steepening at the highest energies due to synchrotron losses. We also expected softer spectra at $z \approx 0$ and harder above the disk ($|z| > 1\, \kpc$) in case B0.5, relative to case B0.0, due to the momentum dependence of the diffusion coefficient.
In test C the adiabatic process is considered together with advective and diffusive transport. The spectrum was expected to reveal a shift towards lower energies due to wind expansion.
In test D all the mentioned processes were included, and we expected to see the combination of all the aforementioned effects.
\subsection{Results}
\begin{figure}
   \newcommand\crwh{19.0cm}
   \newcommand\crwsep{5.pt}
   \centerline{
   \begin{overpic}[height=\crwh,trim={5.725cm 2cm 1cm 0},clip]{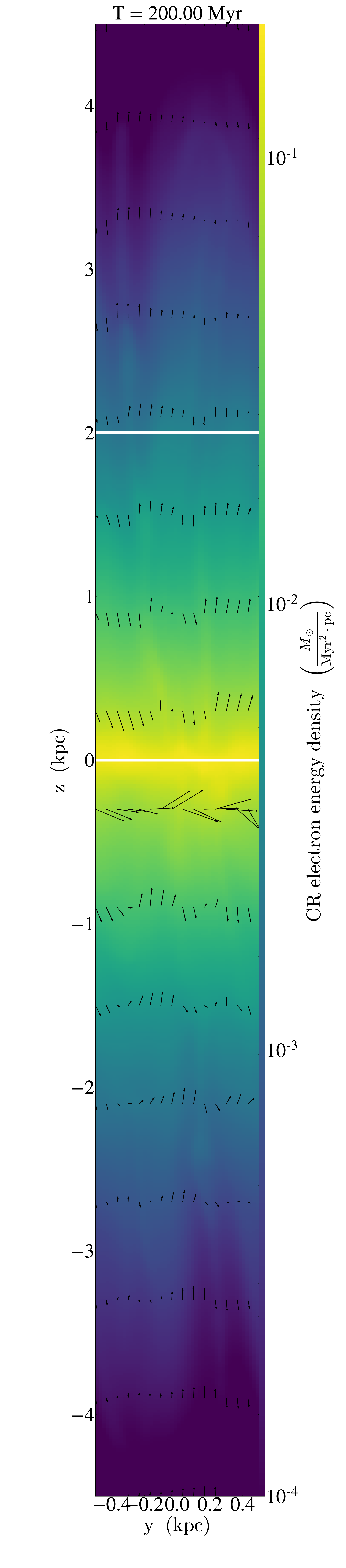}
      \put(3.5,5.){{\footnotesize \color{white}{D0.5: bin 8}}}
   \end{overpic}\hspace{\crwsep}\hspace{\crwsep}
   \begin{overpic}[height=\crwh,trim={11.725cm 2cm 1.5cm 0},clip]{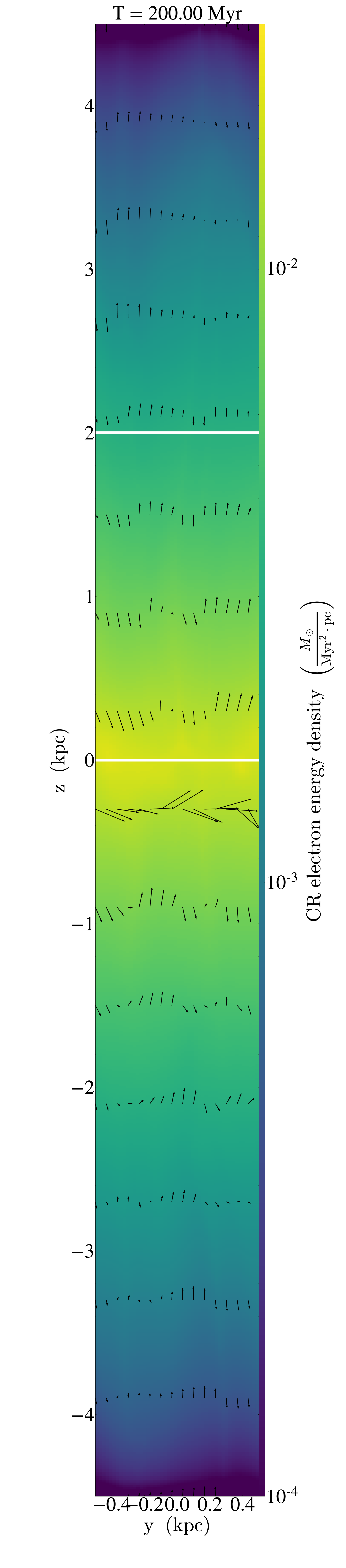}
      \put(0.5,5.){{\footnotesize \color{white}{D0.5: bin 15}}}
   \end{overpic}
   }
   \vspace{-5pt}
   \caption{Distribution of CR electron energy density in physical domain at $t=200$ Myr in test D0.5, for 8-th (left) and 15-th bins (right), encompassing momentum ranges $\p_{8} \in \left[ 215, 464\right)$ and $\p_{15}\in \left[4.65\times10^4, 10^5 \right)$, respectively. White lines  mark the regions from which $z$-averaged spectra were taken. 
   Vectors show magnetic field directions and relative strengths.}
   \label{fig:piernik_wind}
\end{figure}

\begin{figure}[htp]
   \newcommand\specwd{4.5cm}
   \newcommand{\spech}{6.25cm}
   \newcommand\specsh{-4.pt}

   \hspace{-10pt}\includegraphics[width=4.79cm,trim={0.1cm 1.15cm 0.15cm 0},clip]{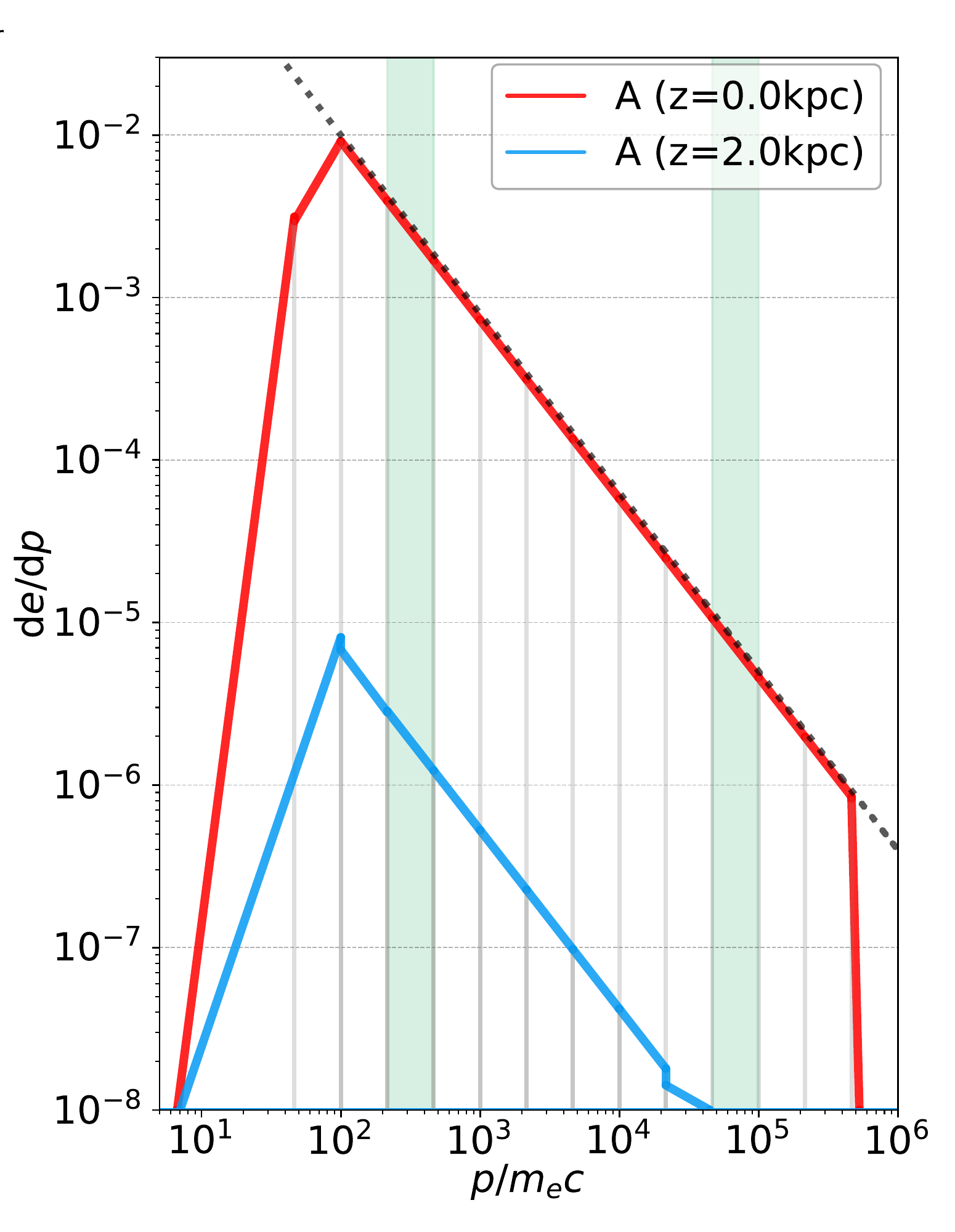}\hspace{\specsh}
   \includegraphics[width=4.13cm,trim={2.4cm 1.15cm 0 0},clip]{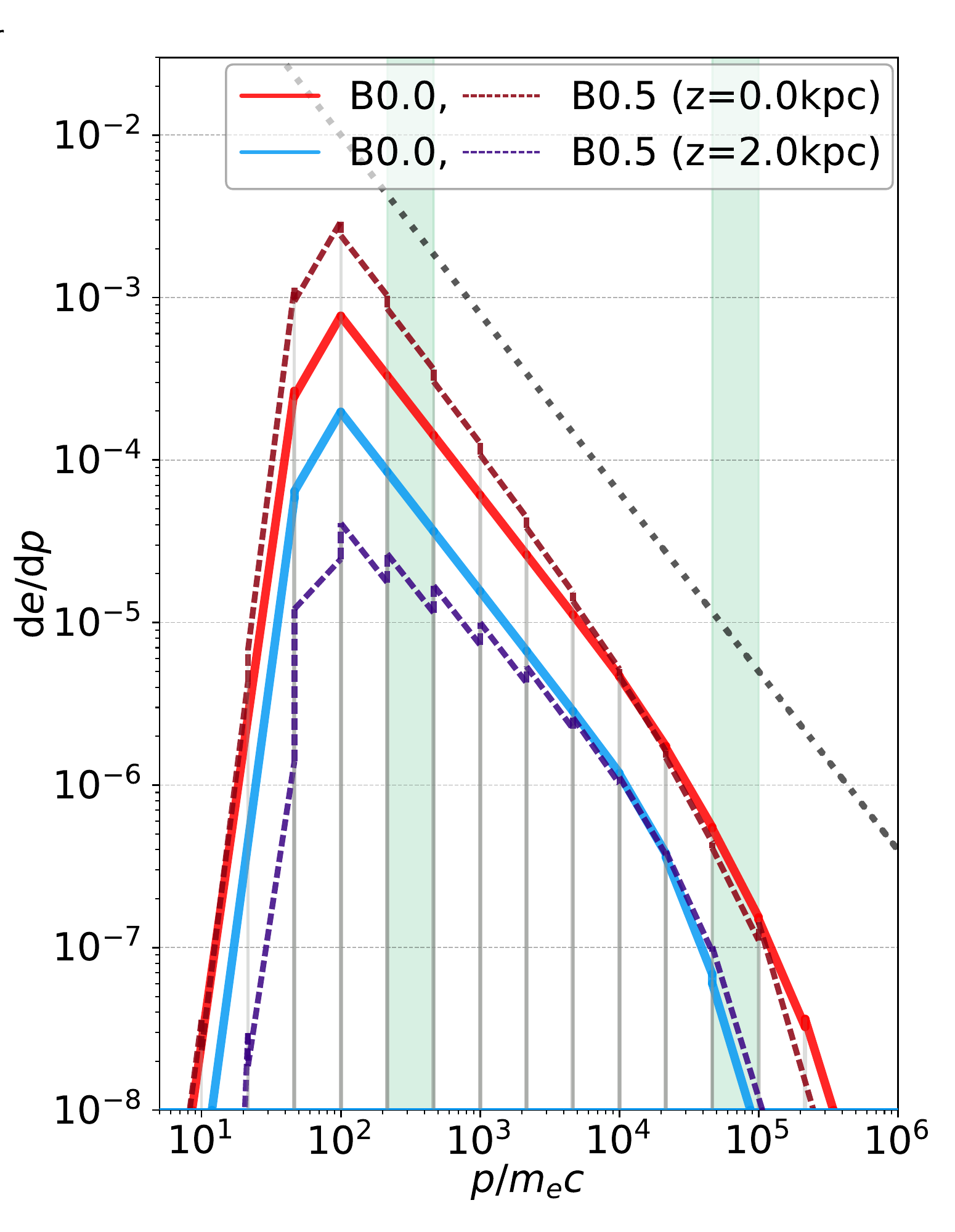}\vspace{-0.175cm}

   \hspace{-10pt}\includegraphics[width=4.79cm,trim={0.1cm 0 0.15cm 0},clip]{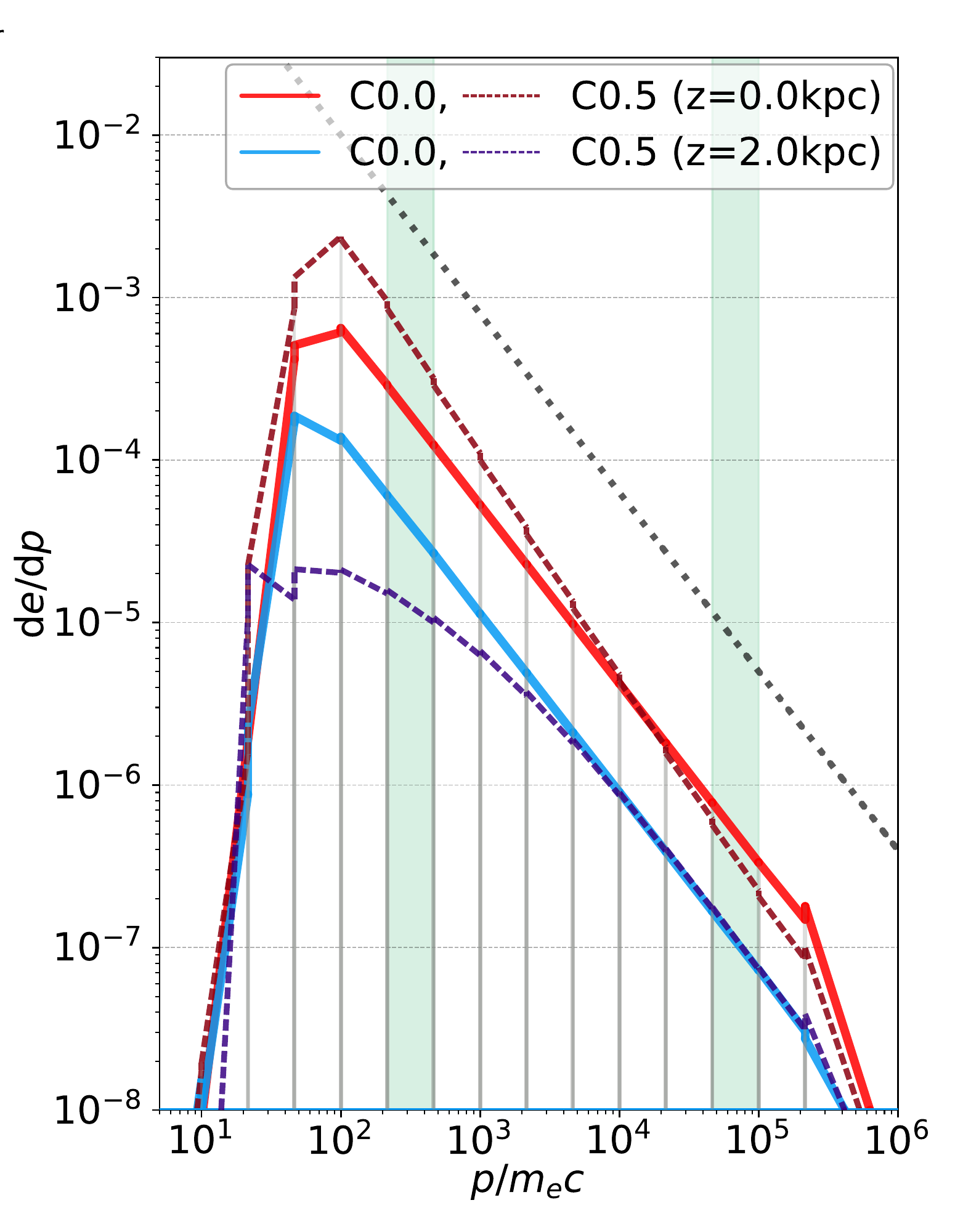}\hspace{-1.75pt}\includegraphics[width=4.13cm,trim={2.4cm 0 0 0},clip]{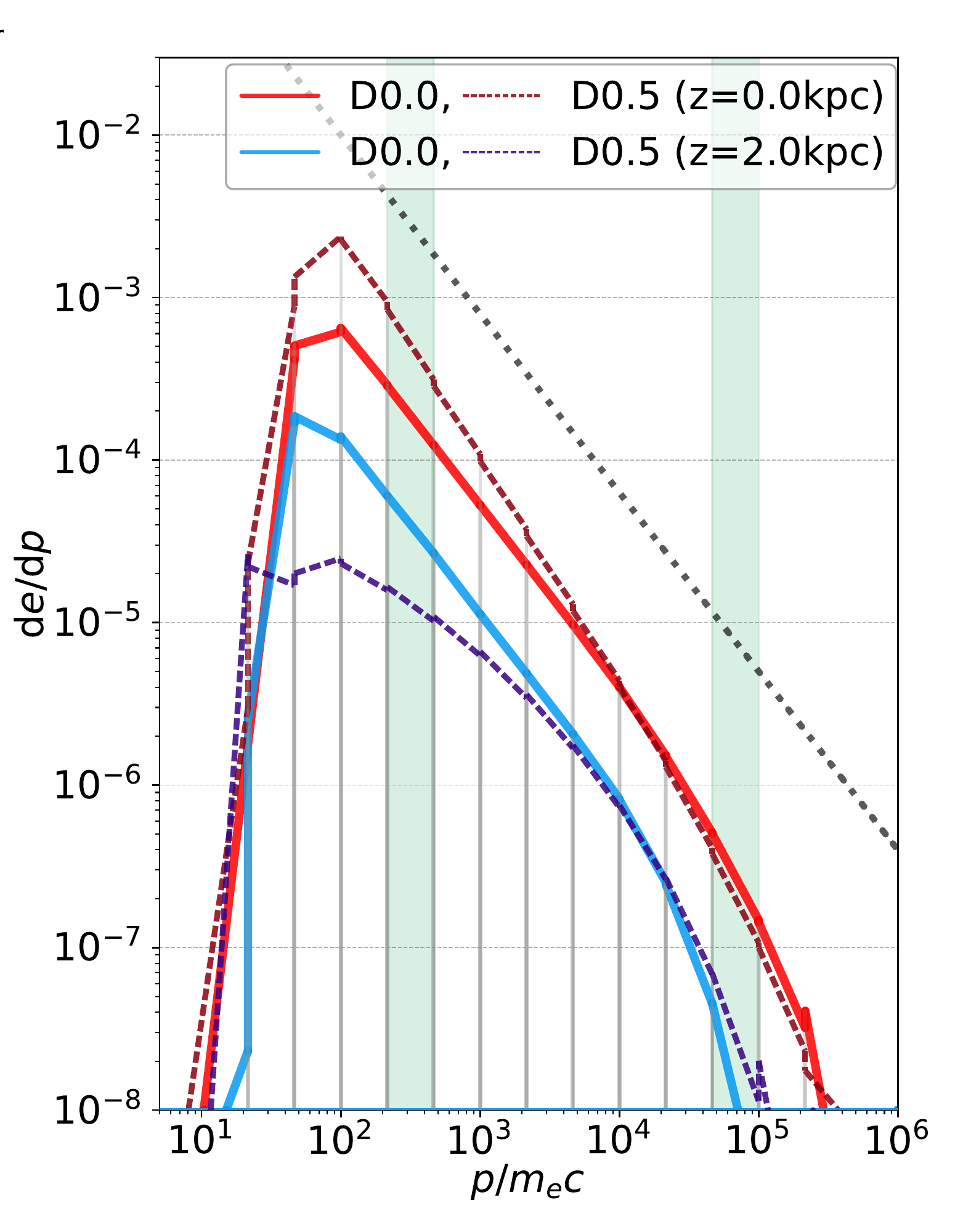}
   \caption{Energy density spectra averaged at two altitudes: $z=0$ (red lines) and $z=2\,\kpc$ (blue lines) for momentum-independent diffusion (solid lines) and momentum dependent diffusion (dashed lines). Test names are indicated at the top of each panel, and their settings are specified in Table \ref{tab:mcrwind-test_names}. Subsequent panels show the effects of different combinations of transport and spectral processes on the spectrum. The dotted gray line marks the initial slope ($q_0 = 4.1$) of the spectrum and marks spectrum amplitude at $z=0\,\kpc$ in test A. Spatial distribution of the energy density in the bins highlighted with the light-green shade, corresponding to $\p_{8} \in \left[ 215, 464\right]$ (8-th momentum bin) and $\p_{15} \in \left[4.65\times10^4, 10^5\right]$ (15-th bin), are presented in Figure~\ref{fig:piernik_wind} for the D0.5 test.}
\label{fig:piernik_wind_spectra}
\end{figure}

In Figure~\ref{fig:piernik_wind} we present domain distribution of CR-electrons for $t \approx 200\,\Myr$ for test D0.5, and in Figure~\ref{fig:piernik_wind_spectra} we show corresponding spectra of CR electrons, averaged over a horizontal plane to eliminate fluctuations resulting from varying local conditions. Positions of taken spectra are marked on Figure~\ref{fig:piernik_wind} with horizontal white lines for the disk plane ($z = 0\, \kpc$) and for an elevated location at $z = 2\, \kpc$. For comparison's sake we chose bins no. 8 and 15, with respective Lorentz factors $\p_{8} \in \left[ 2.15\times10^2,  4.64\times 10^2\right)$ for the 8-th bin, and $\p_{15} \in \left[4.64\times10^4, 10^5 \right)$ for the 15-th bin; these ranges are highlighted in the spectra shown in Figure~\ref{fig:piernik_wind_spectra}. In the momentum-dependent diffusion cases
(i.e., where $\alpha_\kappa = 0.5$) for bins 8 and 15, the ratio of $\kappa_{\parallel,\perp}(p)$ is $\sim 15$.
\\We summarize the test simulation results as follows:
\begin{enumerate}
   \item In the pure advection test A the source spectrum was being injected in random SN explosions and propagated throughout the domain without a change of shape and only with a drop of amplitude at higher altitudes. 
   The only difference appears near the cutoffs of the spectrum, owing to a numerical inaccuracy of the algorithm.

   \item 
   Test B included advective and diffusive transport together with synchrotron cooling. 
   The synchrotron cooling effect steepens the spectra in the high energy range in both cases B0.0 and B0.5, and is apparent beyond $\p \sim 2.15 \times 10^4$, producing a mild break. Diffusion allows particles to travel further from the acceleration sites, resulting in the averaged spectrum amplitudes at 2$\,\kpc$ being higher than in the test A, at the expense of the CR population in the central part.
   The differences in slopes in low and moderate energy ranges are owed to different momentum dependence of diffusion coefficients: in B0.0 case the injection slope is retained, while in case B0.5 spectrum slope steepens with distance, similarly to the results of tests presented in Section \ref{sect:piernik_diffusion}.

   \item Test C included the adiabatic process together with advective and diffusive transport. At $z=0\,\kpc$ in C0.0, we observe no difference in spectral slopes as relative to A0.0 and B0.0. 
   However at higher $z$ 
   we observe adiabatic shift towards lower energies, 
   most prominent for the spectral peak.    This effect results from adiabatic expansion associated with the divergent galactic wind outflow due to the overall CR vertical pressure provided by the CR population.
   In case C0.5 the spectrum above the disk reveals combined effects of momentum-dependent diffusion and adiabatic cooling. Diffusion dominates in the upper part of momentum range, while it is less efficient at the the lower momenta, resulting in smaller spectrum peaks at higher $z$. The fluctuation of the slope near the lower cut-off is an artifact.

   \item  Test D included all the considered processes. 
   The spectra of cases D0.0 and D0.5 show a mild break near $\p\simeq 2.15 \times 10^4$ similar to the one appearing in test B, which can be attributed to the synchrotron cooling  process. The  
   value of high energy cutoff in case D0.5 at $z=2 \,\kpc$ is  
   higher than  
   for D0.0 test, revealing the fact that the higher diffusion coefficient in  
   D0.5 (and B0.5) allows particles to escape faster from the strong mid-plane magnetic field region. The opposite statement is valid for the spectra at $z=0\, \kpc$. Similarly to cases B0.5 and C05,  momentum dependent diffusion softens the spectra in the disk and hardens above the disk. The adiabatic expansion effect shifts the spectrum towards lower momenta similarly as in test C.
\end{enumerate}

\section{Summary and conclusions}\label{sect:conclusions}
We described the CRESP module of PIERNIK MHD code, dedicated to cosmic ray spectrum studies in a galactic context. 
We generalized the two-moment piece-wise power-law method of CR propagation in momentum-space, including fast cooling processes, to the form which is applicable for Eulerian grids in three spatial dimensions. We achieved this goal by providing an operational definition of spectral cut-offs, which can serve as a specific kind of boundary conditions suitable for the transport of the power-law type spectrum of relativistic particles subject to fast cooling processes. The synchrotron emission process naturally tends to evacuate all particles from high energy bins, producing steep spectra near the upper cutoff.

We proposed a novel approach extending the method with specific boundary conditions defined by movable boundaries in momentum space. The construction described in Section~\ref{sect:ext_to_euler} allowed us to estimate the positions of spectral cutoffs solely on the basis of the particle number density and energy density in the outer bins of the spectrum. The  solution enabled binding the algorithms of CR spectral evolution with advection and diffusion in space.

We performed a series of elementary tests of the algorithm to validate its individual elements and their common operation in the case of CR-driven galactic wind setup. The tests have proven the robustness of the algorithm. Despite some apparent (but non-essential) inaccuracies, our algorithm provides results which closely correspond to analytical solutions. We obtained a reasonable evolution of CR spectra in the case of CR-driven wind setup, consistent with expectations concerning the effects of advection, diffusion, adiabatic, and synchrotron cooling of  a CR electron population. Performance tests and other tests, not reported in this paper, indicate that the algorithm is fully suitable for production runs including synchrotron-emitting CR electrons in global galactic-scale MHD simulations on timescales as long as several billion years.

\acknowledgements{This work was supported by the National Science Center under the OPUS grant no. 2015/19/B/ST9/02959 and by the Foundation for Polish Science via the FIRST-TEAM grant no POIR.04.04.00-00-5D21/18-00 (PI: A. Karska). Calculations were carried out at the Academic Computer Center in Gdańsk and on the HYDRA cluster at the Institute of Astronomy of Nicolaus Copernicus University in Toruń. We thank Hubert Siejkowski for the preliminary coding of the algorithm, Philipp Girichidis and Krzysztof Katarzyński for helpful discussions, and Artur Gawryszczak for his invaluable contribution in the development of the PIERNIK code.
We thank  Andy Strong and Thorsten Naab for their encouraging support of this work and the Max Planck Institute for Astrophysics in Garching for their kind hospitality towards MH during early stages of the work on this project.
Post-processing and visualization of the results were carried out using the data analysis and visualization package \yt\, by \cite{2011ApJS..192....9T}}
%

%
\appendix
\label{sect:appendix}
\section{Numerical algorithm for solving CR transport equation for piece-wise power-law distribution function}\label{sect:CR_transport_equations}
In this appendix we summarize the piece-wise power-law method proposed by \citet{2001CoPhC.141...17M} for numerical solving of momentum-dependent CR transport equation (\ref{eq:cr_transp_df}).
We assume an isotropic distribution function.

Let us consider evolution of two moments: number density and energy density of the isotropic distribution function $f(\bm{x}, p)$ of CRs in a section $(p_L, p_R)$ of momentum axis.  The evolution equation for CR number density
\begin{equation}
\nLR \equiv \int_{\pL}^{\pR} 4\pi p^2 f(p) dp
\end{equation}
in  momentum range $(\pL,\pR)$, resulting  from Eqn~(\ref{eq:cr_transp_df}), reads
\begin{equation}
   \pder{\nLR}{t}{} = - \nabla\cdot (\bm{v} \nLR)
      + \nabla \left(  \langle  \kappaLRn \rangle  \nabla \nLR \right)
      + \left[\left(\frac{1}{3}  (\nabla \cdot \bm{v} ) p  + b_l(p)  \right ) 4\pi p^2 f(p)\right ]_{p\limr{L}}^{p\limr{R}} +\QLR(p),
   \label{eq:nLR}
\end{equation}
where $ \langle  \kappaLRn \rangle $ is the momentum-averaged diffusion coefficient of CR particles.
\begin{equation}
   \langle  \kappaLRn \rangle \equiv \frac{ \int_{p\limr{L}}^{p\limr{R}} p^2 \kappa(p) \nabla f(p)  dp}{ \int_{p\limr{L}}^{p\limr{R}}  p^2 \nabla f(p) dp },
   \label{eq:kappa_n}
\end{equation}
The first and second terms on the r.h.s of Eqn~(\ref{eq:nLR}) represent the advection and diffusion transport of CRs.
The third term accounts for evolution of particle momentum due to the adiabatic process and radiative cooling mechanisms (synchrotron, inverse Compton,   ionization, Bremsstrahlung, etc.; see, e.g., \citet{2011hea..book.....L}) which are given by the relation
\begin{equation}
   -\left(\der{p}{t}{} \right) = b\limr{tot}(p) \equiv
   \frac{1}{3}  (\nabla \cdot \bm{u} ) \ p  + \frac{4}{3} \frac{\sigma_T}{m_e^2 c^2} (u_B + u\limr{rad})\, p^2   \equiv u_d \, p   + \ubr \,p^2,
   \label{eq:btot}
\end{equation}
{ where $u_B$ is magnetic field energy density, $u\limr{rad}$ is the energy density of CMB and stellar radiation fields, $\sigma_T$ is the Thomson's cross section, and $m_e$ is the electron mass. The symbols $\ubr$ and $u_d$ are introduced to simplify the notation, where magnetic field energy density together with radiation field energy density are absorbed in symbol  $\ubr$.}
The last  term accounts for CR sources
\begin{eqnarray}
   \QLR \equiv \int_{p\limr{L}}^{p\limr{R}} 4\pi p^2   j(p) dp .
\end{eqnarray}
For $\pL = 0$ and $\pR = \infty$ we obtain the momentum-integrated diffusion-advection equation for CR particle number density
\begin{equation}
   \pder{n\limr{CR}}{t}{} = - \nabla\cdot (\bm{v} \, n\limr{CR} )
   + \nabla \left(  \langle  \kappa_n \rangle  \nabla n\limr{CR} \right) +Q\limr{CR}.
   \label{eq:n_all}
\end{equation}
Similarly, energy density of CR particles
\begin{equation}
   \eLR \equiv \int_{p\limr{L}}^{p\limr{R}} 4\pi p^2 T(p) f(p) dp
   \label{eq:eden}
\end{equation}
is described by the following equation resulting from Eqn~(\ref{eq:cr_transp_df})
\begin{eqnarray}
   \pder{\eLR}{t}{} & = & - \nabla\cdot (\bm{v} \eLR)  + \nabla \left(  \langle  \kappa_e \rangle  \nabla \eLR \right)
   + \left[\left(\frac{1}{3}  (\nabla \cdot \bm{v} ) p  + b_l(p)  \right ) 4\pi p^2  f(p) T(p)\right ]_{p\limr{L}}^{p\limr{R}} \nonumber  \\
   &&     -  \int_{p\limr{L}}^{p\limr{R}}   \left(\frac{1}{3}  (\nabla \cdot \bm{v} ) p  + b_l(p)  \right ) 4\pi p^2  f(p)  \frac{c p }{\sqrt{p^2+m^2 c^2}} dp +\SLR,
   \label{eq:eLR}
\end{eqnarray}
where $T(p) \equiv \sqrt{p^2 c^2 + m^2 c^4} - m c^2$ is the particle kinetic energy and $m$ is the particle rest-mass.   We assume relativistic energy range,  therefore we substitute $T(p) \simeq p c$.
The first two r.h.s terms represent spatial advection and diffusion, respectively, the third one describes advection in the momentum space, and the fourth term represents adiabatic compression/expansion and  radiative losses. The last  term
\begin{eqnarray}
   \SLR \equiv \int_{p\limr{L}}^{p\limr{R}} 4\pi p^2   j(p) T(p) dp
\end{eqnarray}
represents sources of CRs. The momentum--averaged diffusion coefficient of CR energy density is defined as
\begin{equation}
   \langle  \kappa^{LRe} \rangle \equiv \frac{ \int_{\pL}^{\pR} p^2 T(p) \kappa(p) \nabla f(p)  dp}{ \int_{\pL}^{\pR}  p^2 T(p)\nabla f(p) dp } .
   \label{eq:kappa_e}
\end{equation}
The CR pressure contribution from particles in the momentum range $[p\limr{L},p\limr{R}]$  can be computed as
\begin{equation}
   P_{\CR}^\LR\equiv \frac{4\pi}{3}  \int_{p\limr{L}}^{p\limr{R}}  p^3 v(p)f(p) dp {\approx} \; \; \frac{4\pi}{3}  \int_{p\limr{L}}^{p\limr{R}}  p^3 c f(p) dp.
   \label{eq:crs_pressure}
\end{equation}
The latter approximate equality indicates that the expression is valid only in the ultrarelativistic limit ($p \gg mc$).
For $p\limr{L} = 0$, $p\limr{R}=\infty$ and $b_l=0$ we get the momentum-integrated form of the diffusion-advection equation for CR energy density
\begin{equation}
   \pder{\eCR}{t}{} = - \nabla\cdot (\bm{v} \eCR)  + \nabla \left(  \langle  \kappa_e \rangle  \nabla \eCR \right)
   -  \PCR   (\nabla \cdot \bm{v} )  +\SCR(p).
   \label{eq:e_all}
\end{equation}
In the absence of any energy loss processes, the adiabatic condition
$P_{\CR} = (\gamma_{\CR} -1) e_{\CR} $
where $\gamma_{\CR}$ is the adiabatic index of CR gas, can be used to complete the system of equations.  We keep the subscript CR, whenever it is necessary,  to avoid confusion with similar symbols characterizing the thermal gas component.
\subsection{Piece-wise power-law distribution function \label{sect:f_power_law}}
Following \cite{2001CoPhC.141...17M} we assume a piece-wise power-law distribution function defined on a section $[\plmh,\plph]$ of momentum axis. It is convenient to choose  logarithmic scaling for particle bins with fixed bin width equal to $ \Delta w_l = \log_{10} (\plmh/\plph) = \log ( p\limr{max} / p\limr{min} )/\nbin$, where $\nbin$ is the number of bins.
We assume that a piece-wise power-law distribution function characterizes the spectrum of CRs in each cell $(i,j,k)$ of the spatial grid and assign symbol  $\l$ as the index of the actual bin.
For the sake of clarity of notation, we omit the label 'CR' and spatial indexes $(i,j,k))$ in the following considerations.
\begin{equation}
    f(p) = \flmh \left(\frac{p}{\plmh}\right)^{-\ql} \quad \mbox{for}\quad p \in [p_\lmh, p_\lph] \label{eq:dist_fun-eul}
\end{equation}
where the distribution function amplitudes $ f_{\lmh}$ are defined on the left edges of momentum bins $\plmh$ and spectral indices $\ql$ are attributed to bin interiors.
The number density of CR particles in  the $\l$-th bin   is
\begin{equation}
   \nl  =  \int_{\plmh}^{\plph} 4\pi p^2 f(p) dp =\  4\pi \flmh \plmh^3 \times  \left\{
   \begin{array}{c c}
      \frac{1}{{3-\ql}} \left( \left( \frac{\plph}{\plmh} \right)^{3-\ql}-1\right) & \quad \mbox{if} \quad \ql\ne\ 3,  \\
      \log \left(\frac{\plph}{\plmh}\right)                    &\quad \mbox{if} \quad \ql = 3.
   \end{array}
   \right.
   \label{eq:n_l}
\end{equation}
Similarly, the energy density of particles in  the $\l$-th bin given by Eqn~(\ref{eq:eden}) in the relativistic limit is
\begin{equation}
   \el=  \int_{\plmh}^{\plph} 4\pi c p^3 f(p) dp = 4\pi c f_\lmh p^4_\lmh \times \left\{
   \begin{array}{c c}
      \frac{1}{{4-\ql}} \left(\left( \frac{p_\lph}{\plmh} \right)^{4-\ql}-1 \right) & \quad\ \mbox{if} \quad \ql \ne 4,    \\
      \log \left(\frac{p_\lph}{\plmh}\right)                     &\quad \mbox{if} \quad \ql = 4.
   \end{array}
   \right.
   \label{eq:e_l}
\end{equation}
The above expressions differentiate between $\ql = 3$, $\ql =  4$, and other cases.
Number density $\nl$ and energy density $\el$ are stored in a global 4D array for each cell and bin $(l,i,j,k)$, which is updated by the spatial  transport and spectrum evolution algorithms, while the distribution function $f$ and spectral index $q$ are recomputed whenever the spectrum evolution algorithm is invoked.
Evolution of the CR spectrum in momentum space is governed by the third terms of equations (\ref{eq:nLR}) and (\ref{eq:eLR}) including adiabatic process (cooling and heating)  synchrotron, inverse Compton, and other loss mechanisms.  The corresponding evolution equations  for particle number density and energy density on a discrete 1D momentum grid   can be represented as
\begin{equation}
   \nl^{t + \Delta t} = \nl^{t} - \left( \dnulph - \dnulmh  \right),
   \label{eq:ncr_update}
\end{equation}
and
\begin{equation}
   \el^{t + \Delta t} \left(1 + \frac{\Delta t}{2} \Rl^{t+\Delta t} \right)= \el^{t} \left(1 - \frac{\Delta t}{2} \Rl^{t} \right) -  \left( \deulph - \deulmh  \right)
   \label{eq:ecr_update_implicit}
\end{equation}
where the term proportional to $\Rl$ has been integrated implicitly and
$\dnulmh$, $\dnulph $, $\deulmh $ and $\deulph $ are bin interface fluxes of particle concentration and particle energy density integrated over the timestep $\Delta t$, resulting from particle acceleration, adiabatic compression or expansion, and radiative cooling processes.
A simple first order approximation to the solution of energy equation (\ref{eq:eLR})
\begin{equation}
   \el^{t + \Delta t} = \el^{t} -  \left( \deulph - \deulmh  \right) -  \Delta t \Rl^{t} \el^{t},
   \label{eq:ecr_update}
\end{equation}
has been applied in this paper, which  might be upgraded to a higher order solution in future.
The factors $\Rl $ in energy source terms, defined as the energy change rate per unit energy  ($\dot{e}/e$), are computed separately for each bin and cell
\begin{equation}
   \Rl  \equiv \frac{1}{\el} \int_{\plmh}^{\plph} 4\pi c p^3  \flmh (p) b(p) dp = u_d +
   \frac{4 \pi c \,\ubr }{\el} \left\{
    \begin{array}{cc}
      \frac{1}{5-\ql} \left(\left( \frac{\plph}{\plmh}\right )^{5-\ql}-1 \right)&\quad \mbox{if} \quad \ql \ne 5    \\
      \log \left(\frac{\plph}{\plmh}\right)    &\quad \mbox{if} \quad \ql = 5
   \end{array}
   \right.
   \label{eq:e_sink_R}
\end{equation}
It is worth noting that  the evolution equation for particle number density represents a conservative transport of CR particles on the momentum grid, while the energy equation is a conservation law with source terms. Due to heating and cooling processes, particles are only relocated on the momentum grid, and the total number of particles is conserved. A part of the particles' energy is relocated while the other part is radiated away from the system or exchanged with thermal plasma via adiabatic compression or expansion.
To compute the fluxes we first integrate of Eqn~(\ref{eq:btot}) in the time interval $(t, t+\Delta t)$. For the synchrotron and adiabatic processes taken separately we get
\begin{equation}
   \left. p(t+\Delta t)\right|_{\textrm{synch}} = p(t) \left( 1 + \ubr \, p_t  \Delta t \right)^{-1} \quad \mbox{and} \quad
   \left. p(t+\Delta t)\right|_{\textrm{adiab}} = p(t) \exp (-u_d \Delta t).
   \label{eq:p_t+dt}
\end{equation}
with its Taylor expansion up to the third order being
\begin{eqnarray}
   p^{t+\Delta t} & \approx & p^t\left( 1 - \ubr \; p^t \Delta t + \frac{1}{2}(\ubr \; p^t \Delta t)^2 - \frac{1}{6}(\ubr \; p^t\Delta t )^3  \right) \label{eq:synchrotron_cooling_Taylor3}
\end{eqnarray}
Taylor expansion of equation (\ref{eq:btot}) for adiabatic cooling part up to the third order yields:
\begin{eqnarray}
   p^{t+\Delta t} & \approx & p^t\left(1 -\; u_d \Delta t + \, \frac{1}{2}u_d^2 \, \Delta t^2 - \, \frac{1}{6} u_d^3 \, \Delta t^3 \right) \label{eq:adiabatic_cooling_Taylor3}
\end{eqnarray}
Computation  of particle and energy fluxes relies on the analysis of particle flow through interfaces of momentum bins.
We identify particles crossing the interface located at $\plmh$ within the time interval $[t, t+ \Delta t]$.
We define the upstream momentum $\pu$, depending on $\Delta t$, through the requirement that particles located at $\pu$  at time $t$, cross the interface $\plmh$ at $t+\Delta t$. By setting $p(t) = \pulmh$ and $p(t+\Delta t) = \plmh$ we find
\begin{equation}
   \left. \pulmh  \right|_{\textrm{synch}} = \plmh \left( 1 -  \ubr \, \plmh  \Delta t \right)^{-1} \quad \mbox{and} \quad
   \left. \pulmh  \right|_{\textrm{adiab}} =  \plmh \exp (u_d \Delta t)
   \label{eq:p_u}
\end{equation}
for the synchrotron and adiabatic processes, respectively. Number density and energy density fluxes can be computed now by integration of both quantities in the limits $\left[\plmh, \pulmh\right]$ in the case of cooling and $\left[\pulmh, \plmh\right]$ in the case of heating.
The values of upstream momentum $\pulmh$ can be used to compute fluxes of particles crossing bin boundaries within the time interval $\Delta t$. In the case of cooling ($ \pulmh > \plmh$) we get:
\begin{eqnarray}
   \dnulmh  & = & 4 \pi \flmh \plmh^3 \times \left\{
   \begin{array}{c c}
      \;  \frac{1}{3 - \ql}\left(\left(\frac{\pulmh}{\plmh}\right)^{3 - \qlmo } - 1\right) \quad & \quad \mathrm{if} \quad  \ql \neq  3,  \\
      \log{\left( \frac{\pulmh}{\plmh} \right)} \quad &\quad  \mathrm{if} \quad  \ql = 3,
   \end{array}
   \right.
   \label{eq:nupw_l}
\end{eqnarray}
\begin{eqnarray}
   \deulmh  & = & 4 \pi c \flmh \plmh^4 \times \left\{
   \begin{array}{c c}
      \;  \frac{1}{4 - \ql} \left(\left({\frac{\pulmh}{\plmh}}\right)^{4 - \qlmo }- 1\right)\quad & \quad \mathrm{if} \quad  \ql \neq  4,  \\
      \log{\left( \frac{\pulmh}{\plmh} \right)} \quad &\quad  \mathrm{if} \quad  \ql = 4,
   \end{array}
   \right.
   \label{eq:eupw_l}
\end{eqnarray}
and in the case of heating  ($ \pulmh < \plmh$) the analogous formulae are
\begin{eqnarray}
   \dnulmh & = & 4 \pi \flmth \pulmht \left( \frac{p_\lmth} {\pulmh} \right)^{\qlmo} \times \left\{
   \begin{array}{c c}
      \frac{1}{3 - \qlmo}  \left( \left(\frac{p_\lmh}{\pulmh} \right)^{3 - \qlmo } - 1 \right)\quad & \quad \mbox{if} \quad \qlmo \neq 3,  \\
      \log \left(  \frac{p_\lmh}{\pulmh}   \right) \quad &\quad  \mbox{if} \quad  \qlmo =  3,
   \end{array}
   \right.
   \label{eq:nupw_r}
\end{eqnarray}
\begin{eqnarray}
   \deulmh & = & 4 \pi c \flmth {\pulmhf} \left( \frac{p_\lmth} {\pulmh} \right)^{\qlmo} \times \left\{
   \begin{array}{c c}
      \frac{1}{4 - \qlmo}\left( \left( \frac{p_\lmh}{\pulmh} \right)^{4 - \qlmo } - 1 \right) \quad & \quad \mbox{if} \quad \qlmo \neq 4,  \\
      \log \left(  \frac{p_\lmh}{\pulmh}   \right) \quad &\quad  \mbox{if} \quad  \qlmo =  4,
   \end{array}
   \right.
   \label{eq:eupw_r}
\end{eqnarray}
where  $\pulmh $ is  the upstream momentum for the bin interface placed at  $\plmh$.

\subsection{Reconstruction of the distribution function}\label{sect:algorithm_implementation}
In the piece-wise power-law approximation, the whole spectrum of CR particles at each cell of spatial grid can be described equivalently by two sets of discretized quantities:
\begin{eqnarray}
   (n_l, e_l)  \Leftrightarrow (f_{\lmh}, q_l), \quad i=1, \nbin, \nonumber
\end{eqnarray}
The slopes $q_l$ of the piece-wise power-law distribution function can be retrieved by taking
the ratio of particle energy density $e_l$ to the number density $n_l$ given by eqs.~(\ref{eq:n_l}) and (\ref{eq:e_l}):
\begin{equation}
   \frac{e_l}{n_l c \plmh} = \left\{
   \begin{array}{l l}
      \vspace{6pt}   \frac{\left(\frac{p_\lph}{\plmh}\right)^{4 - \ql} - 1 } {\log \left( \frac{p_\lph}{\plmh} \right) \left( 4 - \ql \right) }     & \mathrm{if} \; q = 3  \\
      \vspace{6pt}   \frac{ \log \left( \frac{p_\lph}{\plmh} \right) (3 - \ql) }  {\left(\frac{p_\lph}{\plmh} \right)^{ 3 - \ql} - 1 }    & \mathrm{if} \; q = 4 \\
      \frac{3 - \ql} { 4 - \ql} \frac{\left(\frac{p_\lph}{\plmh} \right) ^ { 4 - \ql} -1 }{ \left( \frac{p_\lph}{\plmh}\right)^{3-\ql} - 1} \, \, , & \mathrm{otherwise}.
   \end{array}
   \right.
   \label{eq:enpc-q}
\end{equation}
Equation (\ref{eq:enpc-q}) can be solved for unknown values of $\ql$ with the aid of a Newton-Raphson-type numerical method.
In the practical implementation we replace the strict condition of $q=3$ by  $|q-3| < \epsilon_q  $ and $q=4$ by  $|q-4| < \epsilon_q  $,  where  $\epsilon_q = 10^{-2}$ seems to be an optimal choice.
If the root-finder attempts to converge to a steep solution  $|q_l| > q_\ind{max}$ then $q_l =  q_\ind{big}$ or $q_l =   -q_\ind{big} $ is used to avoid floating point exceptions in subsequent computations. The value of  $q_\ind{big} $ is  typically set to $30$.
With the computed values of $\ql$ the distribution function amplitudes $\flmh$ can be obtained from~(\ref{eq:n_l}).
\subsubsection{Timestep constraints}
In order to ensure stability of the numerical scheme defined by eqs.~(\ref{eq:ncr_update}-\ref{eq:ecr_update}) we have to restrict the timestep of numerical integration by the condition that particles cannot move more than one bin-width in momentum space
\begin{equation}
   \max\limits_{{1 \leq l \leq \nbin}} \left| \log_{10} \frac{\pulmh}{\plmh} \right|  \leq \cflcrs \;\Delta w
   \label{eq:cfl_crs}
\end{equation}
where $\cflcrs \leq 1$ is a Courant like number controlling the evolution of CR population in momentum space, $\pulmh$ is given by formulae~(\ref{eq:p_u}),
and  $ \Delta w = \log_{10} \left( \frac{p_\lph}{\plmh} \right)$. For practical reasons we use the first order Taylor expansion of ~(\ref{eq:p_u}) in $\Delta t$ and obtain the timestep limit for each individual process
\begin{equation}
   \Delta t_\mathrm{adiab} = \cflcrs  \log_{10} \frac{\Delta w}{| u_d|}, \label{eq:timestep_adiab}
\end{equation}
and
\begin{equation}
   \Delta t_\mathrm{synch} =  \cflcrs \frac{\Delta w}{p\limr{up} \, \ubr}
   \label{eq:timestep_synch}.
\end{equation}
To accurately follow spectrum evolution
in the presence of adiabatic and synchrotron  processes we impose  $\cflcrs \simeq 0.1-0.2 $.

\bibliographystyle{aasjournal}
\begin{scriptsize}
   \bibliography{bibliography_list}
\end{scriptsize}

\end{document}